# USING BAYESIAN NETWORK ANALYSIS TO REVEAL COMPLEX NATURES OF RELATIONSHIPS

**Panchika Lortaraprasert[1], Pongpak Manoret[1], Chanati Jantrachotechatchawan[2], Kobchai Duangrattanalert[3,*]**

[1] Triam Udom Suksa School, Thailand;
puiphai01@gmail.com (P.L.); pongpak282546@gmail.com (P.M.)
[2] Research Division, Faculty of Medicine Siriraj Hospital, Mahidol University, Thailand;
chanati.jan@mahidol.ac.th (C.J.)
[3] Chulalongkorn University Technology Center (UTC), Chulalongkorn University Thailand;
kobchai.d@chula.ac.th (K.D.)

* Corresponding author

---

## ABSTRACT

Relationships are vital for mankind in many aspects. According to Maslow hierarchy of needs, it is suggested that while a healthy relationship is an essential part of a human life that fundamentally determines our goals and purposes, an unsuccessful relationship can lead to suicide and other major psychological problems. However, a complete understanding of this topic still remains a challenge and the divorce rate is rising more than ever before to almost 50 percents. The objective of this research is to explore the association between each group of behaviors by performing Bayesian network analysis on a large publically available Experiences in Close Relationships Scale, a test of attachment style survey (ECR) data from *openpsychometrics* database. The resulting directed acyclic graph has 2 root nodes (Q02 from avoidant and Q05 from anxious attachment) and 5 end nodes (Q16, Q34, and Q36 from anxious attachment). The network can be divided into 5 clusters, 2 avoidance and 3 anxiety clusters. Furthermore, our list of items in the clusters are consistent with the findings of previous factor analysis studies and our estimated coefficients are significantly correlated with those of one partial correlation network study.

## 1. INTRODUCTION

### 1.1. Background

In the past decades, many countries, including Sweden, Finland, and France, have encountered over 50% of divorce rates [1] [2] [3]. Meanwhile, judicial separation ratios in more than 35 countries surpass 40%. [1] [2] [3] [4] [5] [6] [7] In various aspects, these are serious situations that could lead to major social problems. A study using Maslow's hierarchy of needs-based questionnaire [8] found that love and belonging are needed for maintaining self-esteem and self-actualization. Consequently, intimate relationships can be considered 'psychological needs' such that their subsequent losses are potentially detrimental. Moreover, research from California University also shows significantly increasing chances of suicide in divorce and separated men [1].

The used data in this research are derived from *Experiences in Close Relationships Scale, a test of attachment style (ECR)*. The survey was created in 1998 by Kelly Brannan, Catherine Clake and Phillip Shaver [9]. ECR consists of 36 questions to let users rank from 1-5 in duration of 4 minutes. The majority of test subjects are women, American citizens and people between the ages of 18-30. Classifying individuals based on their anxious or avoidant attachment styles is the main objective of using ECR. However, studies on the interaction or causal influences between the states of relationship have been lacking.

In order to reveal natures of relationships, we decided to use the Bayesian Network to generate the 'close relationship map' which will illustrate states in romantic relationships represented as the 36 items of the ECR that humans could possibly get through. Consequently, the map will not only be able to indicate paths that could solve critical issues, but also capable of interpreting many features, such as signs of unhealthy relationship, cause of relationship impediments, causal interactions between states, and transition points of one attachment style to another.

# 2. RELATED WORK

## 2.1. Effects of close relationship

A range of effects in behaviors from specific types of relationships is identified in recent studies which could be as

| Relationship types | Effect |
|---|---|
| Unhealthy relationship [10] | Poor well-being |
| Relationship with offender after the crime [11] | Slightly higher dependence |
| Full with closeness relationship [12] | Higher rate of criminal protection |
| Full with satisfaction relationship [13] | Higher self-regulation |
| Full with desire to continue online relationship [14] | More cooperative management style |

**Table 1.** Summary of effects of close relationship on other living aspects

## 2.2. Factors which may influence quality of relationship

Similarly, a large number of research focusing on finding elements for healthy relationships are conducted. To illustrate, in 2020, there was a collaboration between the University of Michigan and Eastern Michigan University [15]. In this experiment, they created online surveys for 592 participants over a 12 months duration. For the results, firstly, the research indicated that older people tend to have more positive relationships. Moreover, if elders have negative relationships, their well-being will be less affected compared to other generations.

Furthermore, an experiment from the University of Adelaide with Radboud University provides information about the association between closeness and forgiveness [16]. According to the outcome, there is going to be a positive association when post-transgression is increasing.

Interestingly, while anxious attachment is associated with a higher risk of suicide in elderly [17], avoidant attachment is more negatively correlated with relationship quality in young adults [18].

## 2.3. Factor and network analysis on the ECR and its derivatives

Kelly Brennan, Catherine Clake and Phillip Shaver published the ECR survey in 1998 [9]. Users have to complete 36 questions by ranking each question from 1 (strongly disagree) to 5 (strongly agree) in 4 minutes. These 36 questions are divided equally into items of 2 statistically independent attachment dimensions: avoidant (odd number) and anxious (even number) attachment as described in the attachment theory originated by John Bowlby [19] and Mary Ainsworth [20]. The data was collected from more than 1,000 undergraduate students and these 36 items were selected from a pool of 323 items based on their absolute correlation coefficients with the 2 attachment dimensions. The original purpose of the questionnaire is to classify individuals as one of the following attachment styles (secure, preoccupied, dismissing, and fearful) as shown in **Table 2**.

| | | **Avoidance** | |
|---|---|---|---|
| | | Low | High |
| **Anxiety** | Low | **Secure** | **Dismissing** |
| | High | **Preoccupied** | **Fearful** |

**Table 2**. Attachment styles from combinations of attachment dimensions

Wei et al. (2007) [21] developed a short-form of ECR by using a combination of both statistical and rational criteria to select 12 representative items, 6 from each attachment dimension. They performed exploratory factor analyses separately for the 2 sets or subscales, namely avoidance and anxiety, of items. For each set, principal axis factoring was conducted to extract factors that have eigenvalue higher than 1 and consequently account for the most common variance of the set. The 2 avoidance and 3 anxiety factors are listed in **Table S1**.

Lo et al. (2009) [22] modified the questionnaire statements to also assess the relationship with close others in addition to romantic partners. They studied the responses from patients with metastatic gastrointestinal and lung cancer with a mean age of approximately 60 years old. The data was collected 2 times: Time 1 and Time 2 data were from 309 and 120 participants respectively. A total of 97 participants completed the survey at both times. Factor analysis was conducted on the avoidance and anxiety items together and the 4 identified major factors could be attributed to the following attachment characteristics: F1 for *Frustration about unavailability*, F2 for *Discomfort with closeness*, F3 for *Turning away from others*, and F4 for *Worrying about relationships* (**Table S2**).

Guzmán-González et al. (2019) [23] also performed factor analyses with the 2 attachment dimensions as the 2 factors on a cohort of 943 Chilean adults in order to construct a brief 12-item Spanish version of the ECR. The 2 factors together explained 36.43% of the shared variance. Although most items exhibit factor values corresponding to their intended dimensions (odd - avoidant, even - anxious attachment), there are few exceptions (**Table S3**).

McWilliams and Fried (2019) [24] performed a correlation network analysis on the 10-item Relationship Structure Questionnaire (RSQ) which was derived from the revised ECR by Fraley et al. [25]. The study was conducted in 2 rounds with sample sizes of 310 and 3,710 respectively. Respondents completed 4 copies of the test for their relationship with the following 4 figures: mother, father, romantic partner, and best friend. The Gaussian Graphical Model was applied to the Pearson correlation matrix of the items to estimate the regularized partial correlations among the items. False-positive connections were minimized by reducing very small coefficients to zeros with the graphical least absolute shrinkage and selection operator or GLASSO. The partial correlation coefficients of the romantic partner data from the second study can be viewed in **Table S10A**.

## 2.4. Bayesian Networks usage

Bayesian Network (BN) was invented by Pearl and Kim in the 1980s [26]. It is a graphical model which uses probabilistic distribution instead of deterministic information to estimate the probability of one event given another. In this way, the BN represents the data as a directed acyclic graph (DAG) and models the conditional dependencies between each node and its corresponding parent nodes set of parent nodes and their corresponding child (parent and child). Expectedly, a BN graph still possesses common graph characteristics that can be quantified including different centrality values i.e. these data could be illustrated in various forms. For instance, degree in, degree out, direction, strength, betweenness and closeness.

The application of BN has been observed and proved beneficial tois in many industries. In 2007, customer satisfaction surveys were analysed and an effective approach was suggested introduced by the BN [27]. Likewise, railway transport also used the BN model to examine customer satisfaction by 2016 for Queensland Rail in Australia [28]. Bayesian network has been implemented to show that avoidant attachment of nurses directly influences negative representation of caregiving and that anxious attachment negatively influences self-efficacy and induces emotional exhaustion [29].

To the best of our knowledge, this is the first study that uses Bayesian network analysis to model the causal relationships between the items of attachment style questionnaires including the ECR and its modified variants.

# 3. METHODOLOGY

## 3.1. Data preprocessing

Experiences in Close Relationships Scale, a test of attachment style (ECR) survey is conducted in an online platform called "Open-Source Psychometrics Project' (https://openpsychometrics.org). The website has been providing various types of up-to-date interactive tests and millions of results since 2011. In this case, besides the fact that it was downloaded in April, 2021, used data for this project are from 41,773 participants who have a wide range of backgrounds; ie., age (18-60), homeland (5 regions; Africa, America, Asia, Europe, Oceania).

In the sample selection procedure, '*countrycode*' R package was used for grouping countries, which are in the form of country coding (e.g. US), by their respective continental regions. Hence, the 41,773 samples are from 6 continental regions (Africa, North America, South America, Asia, Europe and Oceania), 2 genders (female, male) and 4 groups of age (18 - 20, 21 - 30, 31 - 40 and 41 - 60). The majority of test subjects are North or South American, women, and people between the ages of 18 - 20 (**Table 3**)

| Type | Group | Quantity |
|---|---|---|
| **Region** | America (North and South) | 25,274 |
| | Europe | 9,401 |
| | Asia | 4,006 |
| | Oceania | 2,390 |
| | Africa | 702 |
| **Gender** | Female | 28,685 |
| | Male | 13,088 |
| **Age** | 18 - 20 | 19,151 |
| | 21 - 30 | 17,718 |
| | 31 - 40 | 7,187 |
| | 41 - 60 | 6,717 |

**Table 3**. Demographic profiles of the online ECR participants.

## 3.2. Structure learning

The data was analyzed by *bnlearn*, a continuously developed R package designed for Bayesian networks (BN) graphical structure learning. There are multiple algorithms to learn an optimal graph structure by a given type of metrics. For score-based structure learning algorithms, a null graph is initialized and directed edges or arcs are added, removed, or reoriented according to a global graph score, which indicates the likelihood of the observed data given the graph. The *tabu search*, an algorithm used in this study, is a score-based algorithm that, unlike other local search algorithms which may be trapped in a local optimal, can find the global optimal with a more flexible procedure but also requires a longer run time. The details on theory behind the BN and usage of *bnlearn* can be read further here [30].

The BN model was trained using *boot.strength()* function that performs bootstrapping with an *m* given sampling size for an *R* number of epochs and provides a strength estimate of each arc/connection as its empirical frequency over a set of *R* networks. The resulting *R* replicates of BN models are then averaged by *averaged.network()* function to yield a final model . In this study, the sampling size *m* is always at 1000. To determine the number of training epochs sufficient for the BN model to capture the structure of this data, we conducted the bootstrapping 5 times for different numbers of epochs (50, 100, 200, 500, 1000, 1500, 3000, and 5000) and record directed and undirected connections counts. The directed (**Figure S2A**) and undirected (**Figure S2B**) connections counts become stable at approximately 123 and at 0 respectively with their standard deviations of less than 1 and 0 when the model is trained for 3000 and 5000 epochs. Thus, training the model for 3000 replicates is a sufficient amount.

## 3.3. Parameter estimation and subgraph clustering

The method that we cooperated in this stage is Maximum likelihood estimation (MLE) which is a method that maximizes the likelihood function to estimate a probability distribution parameter. '*bn.fit*' is the function that we used with a purpose to fit Bayesian network parameters on its given structure according to the dataset. Subsequently, the standard deviation of the residuals is approximately 1.

Densely connected subgraphs or communities were identified via random walks algorithm using the function *cluster_walktrap* from *igraph* R package. Note that from here onward, we will refer to a densely connected subgraph or community as a cluster in order to avoid a connotation of community as a social unit, especially since we are modeling the states of close relationships.

## 3.4. Centrality measures

Four centrality values were computed for the network: degree in, degree out, betweenness, and PageRank. Degree in and degree out are simply the number of incoming edges to and outgoing edges from each node. In the context of a Bayesian network directed graph, the degree in of a given node $v$ is equal to the number of nodes that influence $v$ (parent nodes) and the degree out of $v$ is the number of nodes influenced by $v$. Betweenness centrality of node $v$ measures the ratio of shortest paths between any pairs of nodes that pass the node $v$ over the total number of shortest paths between that given pair of nodes. For a weighted graph, the function *betweenness* from the *igraph* R package automatically interprets weight as distance and the path length is the sum of weights of the edges,

$$C_B(i) = \sum_{u \neq v \neq w} \frac{p_{uw}(v)}{p_{uw}} \quad (1)$$

where $P_{uw}$ is the total number of paths from $u$ to $w$ and $P_{uw}$ is the number of paths from $u$ to $w$ that pass through $v$. However, opposite to the notion of the distance, a coefficient reflects a causal influence or a conditional probability of one node given its parent node. The higher the coefficients are, the higher connectivity or proximity their corresponding edges represent. Therefore, betweenness was calculated using an inverse of coefficients as weights instead.

Google PageRank centrality (PR) scales the contribution from the parents of node $v$ to the PR value of $v$ by the weighted degree of those parent nodes using the concepts of solving an equation for an eigenvector as follows

$$C_{PR} = D(D - aA)^{-1} \quad (2)$$

where $D$ is a diagonal matrix of weighted degrees out with diagonal as maximums of its degree out or 1 to ensure that its determinant is not zero and $D$ is invertible.

## 3.5. Indirect influence calculation

As contributions are estimated as coefficients of a linear system with a general form,

$$x_v = \sum_{v_j \in pa(v)} c_{v_j,v} x_{v_j} + b_v \quad (3)$$

where $pa(v)$ is the set of parent nodes $v_j$ of $v$; and $c_{v_j,v}$ corresponds to a contribution coefficient from $v_j$ to $v$ and a directed edge or arc $e(v_j, v)$.

A change in any node $w$ with respect to node $v_0$, given $P_{v_0w}$ a set of directed paths (sequences of distinct arcs and vertices) from $v_0$ to $w$, can be calculated as a sum of coefficient products along the paths based on the chain rule of derivative as follows

$$\frac{dx_w}{dx_{v_0}} = \sum_{P_{v_0w}} c_{v_m,w}\, c_{v_{m-1},v_{pm}} \cdot\cdot\, c_{v_{i-1},v_{pi}} \cdot\cdot\, c_{v_1,v_{p2}}\, c_{v_0,v_1} \quad (4)$$

Where $v_m \in pa(w),\ v_{m-1} \in pa(v_m), \ldots,\ v_{i-1} \in pa(v_i), \ldots,\ v_0 \in pa(v_1)$ and each path is a sequence $(e(v_0, v_1),\ \ldots,\ e(v_{i-1}, v_i), \ldots,\ e(v_m, w))$. Note that different paths may have different numbers of arcs. A demonstrating example of this concept is in the Supplementary method with a diagram of the network example in **Figure S1**.

## 3.6. Clustering of data from the previous factor analysis studies

To study the distribution and aggregation of all 36 items across the factors of the previous studies, *k*-means clustering was used to identify 2 avoidance and 3 anxiety clusters on the data from Wei et al. [21] and 5 clusters on the data from Lo et al. [22] and Guzmán-González et al. [23]. Function *kmeans* from library *cluster* was run on a given dataset over a range of random seeds from 1 to 4000 to determine which seed corresponds to the initial set of centers that give the smallest *tot.withinss* or total with-in cluster sum of squares.

Clusters were then visualized by *fviz_cluster* from library *factoextra*. For any dataset with 3 or more axes, this function automatically performs principal component analysis (PCA) and plots the data on the first two principal components. Note that while PCA after factor analysis may preserve linear relationships of the original data among the items, the percentages of variance explained by the new PCs were computed for the factors -- not the original data analyzed by the previous studies. A concentration ellipse was constructed to portray the *t*-distribution probability of data points in each cluster at 95% confidence. In other words, items outside an ellipse of a given cluster have probability of belonging to that cluster by chance alone less than 0.05.

## 3.7. Correlation to compare with the previous network study

Correlation analysis between the influence coefficients estimated by our BN model and the partial correlation coefficients inferred by McWilliams and Fried network model was performed to assess the similarity between the two studies. Correlation coefficients were computed twice for the following:

1) the set union of item pairs from both studies such that a coefficient of any pairs not in the intersection between the two sets of item pairs were assumed to be 0,

2) the set intersection of item pairs from both studies.

The t-score was calculated from the Pearson correlation coefficient and the degree of freedom (the number of pairs - 2) as the equation (5) below,

$$t = r\sqrt{\frac{df}{1-r^2}} \quad (5)$$

A *p*-value is then calculated as a two-tailed *t*-distribution given the *t*-score and the degree of freedom.

# 4. RESULTS

## 4.1. People are optimistic in the primary stage of a relationship.

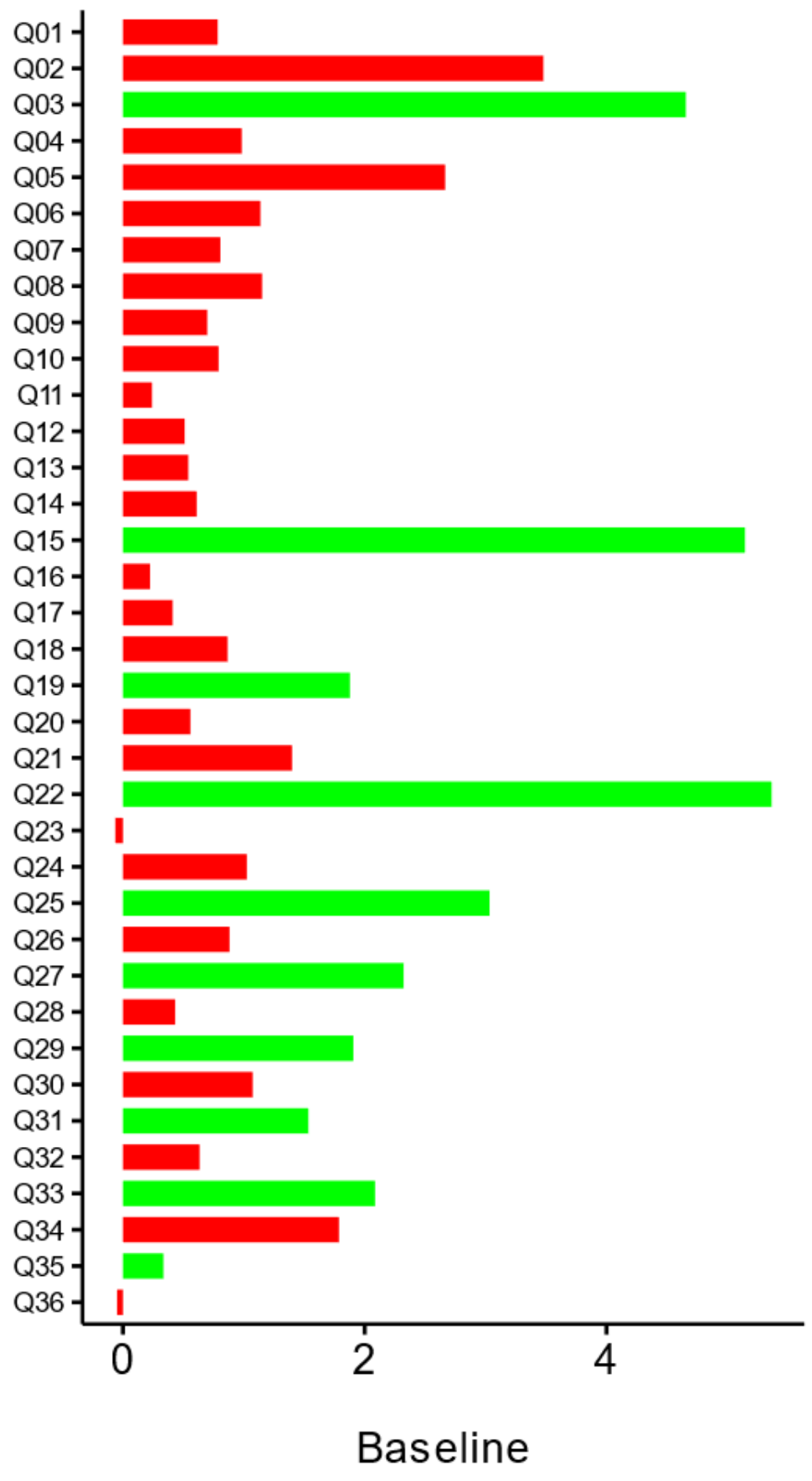

**Figure 1.** Intercepts or baselines of each ECR item estimated from Bayesian network analysis.

Full statements and intercepts or baseline values of all items in the avoidant and anxious attachment groups are listed in **Table S4** and **Table S5** respectively. Intercepts of all 36 nodes are presented in a bar chart (**Figure 1)**. The green color represents a positive behavior or a reversed keyed item and red represents the opposite. It is noticeable that the majority of high baseline questions are positive attachment states and the difference in intercepts between the positive and negative items is statistically significant ($p$ = 0.0003, Mann-Whitney U). For instance, the item Q22 (5.365) is 'I do not often worry about being abandoned. On the other hand, the questions that have low baseline are usually classified as negative behavior. For example, Q16 (0.22), my desire to be very close sometimes scares people away. Interestingly, the two highest baseline negative behavior nodes -- Q02 and Q05 -- are also the two root nodes of the network. Likewise, the second-highest baseline node Q15 is the root node of C5. Thus, it might be concluded that, without any outer influence, while people tend to be optimistic about their relationship, they may also manifest the early signs of anxiety or avoidance.

## 4.2. Characteristic of each cluster

The relationship network is divided into 5 subgraphs: Maintaining distance (C1, m = 10), Worrying of being abandoned (C2, m = 7), Feeling insecure (C3, m = 7), Overly attached (C4, m = 4) and Comfortable opening up or relying on (C5, m = 8)

Clusters C1 and C5 are groups of avoidance questions (**Figure 2**). It can be interpreted that the cluster of distance (C1) would end as suppression of positive state (Q03; *I am very comfortable being close to romantic partners*) or influence/contribution to negative state (Q21; *I find it difficult to allow myself to depend on a romantic partner*). The root node of this cluster is Q05 (*Just when my partner gets close to me I find myself pulling away*). In a cluster of comfortable opening up or relying on (C5) (**Figure 2**), all questions reflect a character for healthy relationships. It may start with a deep conversation (Q15; *I feel comfortable sharing my private thoughts and feelings with my partner.*) and end with comfortable dependence (Q29; *I feel comfortable depending on romantic partners*). All coefficients within the avoidance clusters are listed in **Table S6**. Additionally, the last nodes of these avoidance clusters are not the final nodes of the relationship network.

The remaining clusters C2, C3 and C4 are clusters of anxiety (**Figure 3**). For C2 , the root state is worrying about being abandoned (Q02; *I worry about being abandoned.*). Similar to C1, the last stages of C2 could be either inhibition of positive Q22; I do not often worry about being abandoned) or promotion of negative Q10 (*I often wish that my partner's feelings are as strong as my feelings for him/her*) (**Table S7**). Secondly, from **Figure 3B**, C3 (Feeling insecure) starts with reassurance needed (Q18; *I need a lot of reassurance that I am loved by my partner*) and finish with negative behavior (Q34; *When romantic partner disapproves of me, I feel really bad about myself* or Q36; *I resent it when my partner spends time away from me.*) Lastly, the beginning point of the overly attached cluster (C4) is Q26: *I find that my partner(s) don't want to get as close as I would like*. Then, the last point is Q16: *My desire to be very close sometimes scares people away.*

## 4.3. Relationship between each cluster and overall network characteristics

The connections between each cluster are illustrated in **Figure 4**. All clusters except C2 can both be influenced by or influence another cluster in the network. C2 is not influenced by other clusters and Q02 the root node of C2 is also a root node of the relationship network. Although C1 is influenced by C2, Q05 the root node of C1 is not influenced by any other nodes of other clusters (**Table S8**) and therefore is the second root node of the entire network.

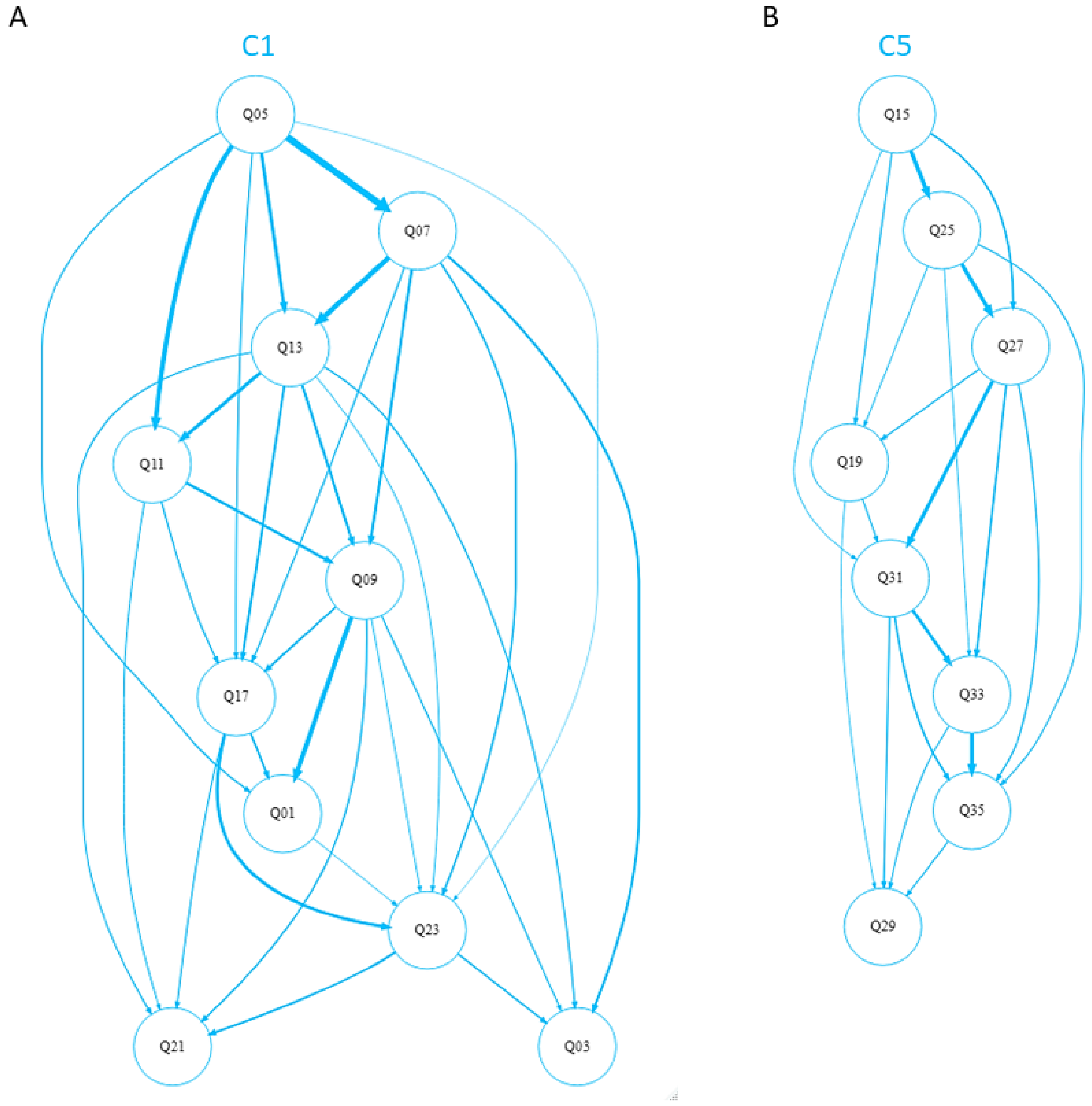

**Figure 2.** Two clusters of the avoidant attachment dimension. Maintaining personal space (C1) subgraph (A), and Comfortable opening up or relying on (C5) subgraph (B).

The strongest connections between clusters are from being not comfortable to open or be close with (C1) to being comfortable to open up (C5), both of which are clusters of avoidance attachment style. Consistently, the clusters of anxiety attachment -- C2, C3, and C4 -- have high to moderate connections to one another as the sums of absolute coefficients are greater than the median of all coefficients at 0.17217 (**Figure S3**). Overall, connections between clusters of the same attachment are stronger than those between clusters of different attachment styles. Q16 from C3 and Q34 and Q36 from C4 do not contribute to any other nodes and hence are the end nodes of the entire network as well.

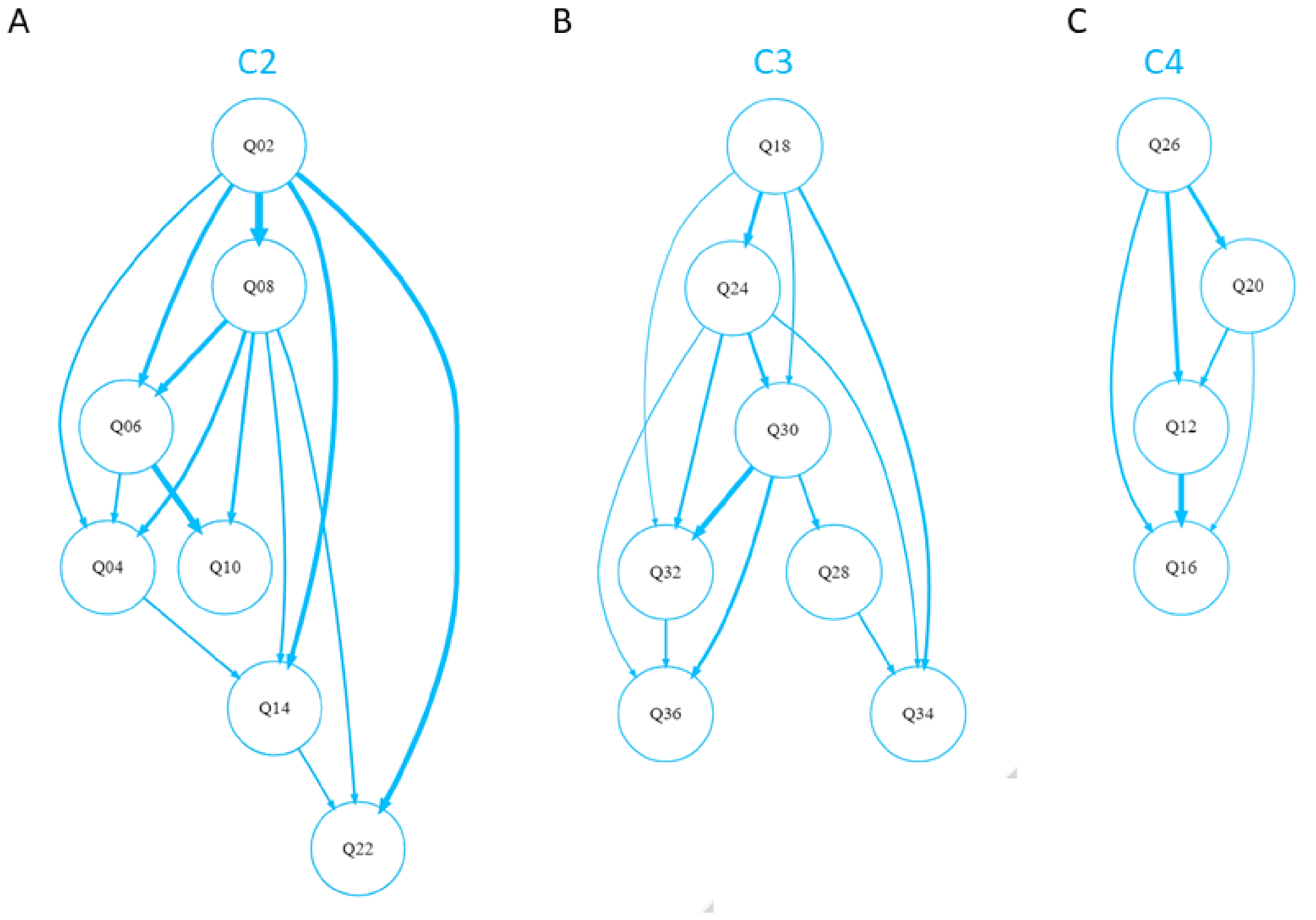

**Figure 3.** Three clusters of the anxious attachment dimension. Worrying of being abandoned (C2) subgraph (A), Feeling insecure (C3) subgraph (B), and Overly attached (C4) subgraph (C).

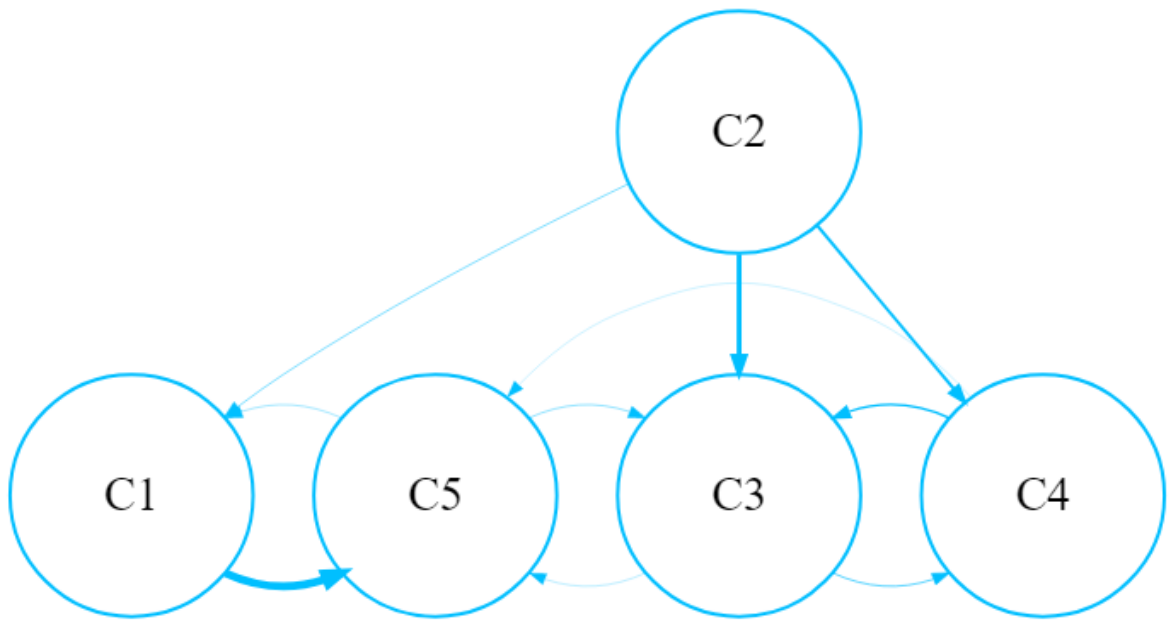

**Figure 4.** Relationship between each subgraph cluster.

The relationship network can be characterized by various centrality values. In terms of degree or the number of connections, Q09 (*I don't feel comfortable opening up to romantic partners*) has the highest number of arcs to other nodes (degree out) (**Figure S4A**) while Q23 (*I prefer not to be too close to romantic partners*) receives the highest number of contributions (degree in) (**Figure S4B**). When inverse absolute coefficients were included as weights or distances of the graph, Q09 and Q23 also have the highest betweenness centrality (**Figure 5**). This can be explained by Q23 receiving influences from Q22 the end node of C2 and also directly contributing to Q03 that in turn provides contributions to 3 other nodes of C5 (**Table S8**). Q09 influences Q03, Q23, and also the  Q09 and Q23 are therefore possibly two important intermediate states, of which intervention may considerably affect the entire relationship network.

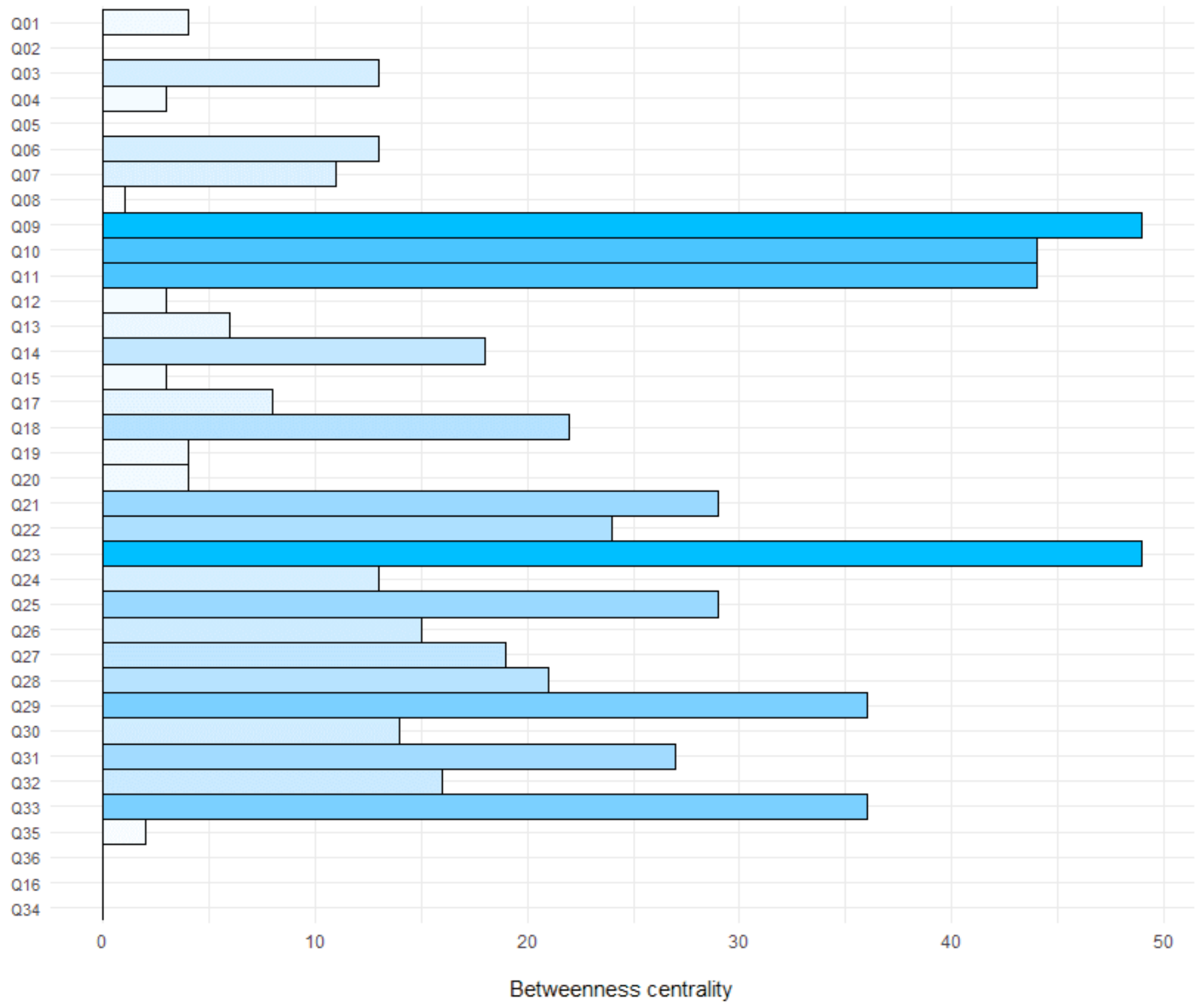

**Figure 5.** Betweenness centrality of all 36 nodes of the relationship network. The intensity of the blue color of each bar is directly correlated with the centrality value.

Google PageRank (PR) centrality evaluates the importance of each node by the input connection weights -- here contribution coefficients -- that have been adjusted for coefficients of other contributions by their corresponding parent nodes. In this relationship network, Q28, Q29, and Q34 have by far the highest PR centrality (**Figure 6**), even above two folds of the fourth-highest and fifth-highest PR values of Q35 and Q36 respectively. This indicates that Q28, Q29, and Q34 are the most important states of the relationship as they relatively receive the most overall contributions from the system.

At first glance, PR calculation seemingly conflicts with the concept of the BN. PR value from a parent node when transferred to its target node is divided by the total number of output edges of that parent node. This is in contrast with the BN model estimation that an influence from a parent node to a particular target node is not affected by the influence from that parent node to the other nodes *per se*. Nevertheless, the purpose of the PR algorithm here is to assess the relative importance of each node compared with others, which does not have to be equivalent to the estimated score or probability of that node. To illustrate, Q28 has the highest PR value despite having a relatively small sum of influence coefficients to Q28 likely because Q28 is the sole target node of Q29. Similarly, Q34 is a sole target node of Q28 as well.

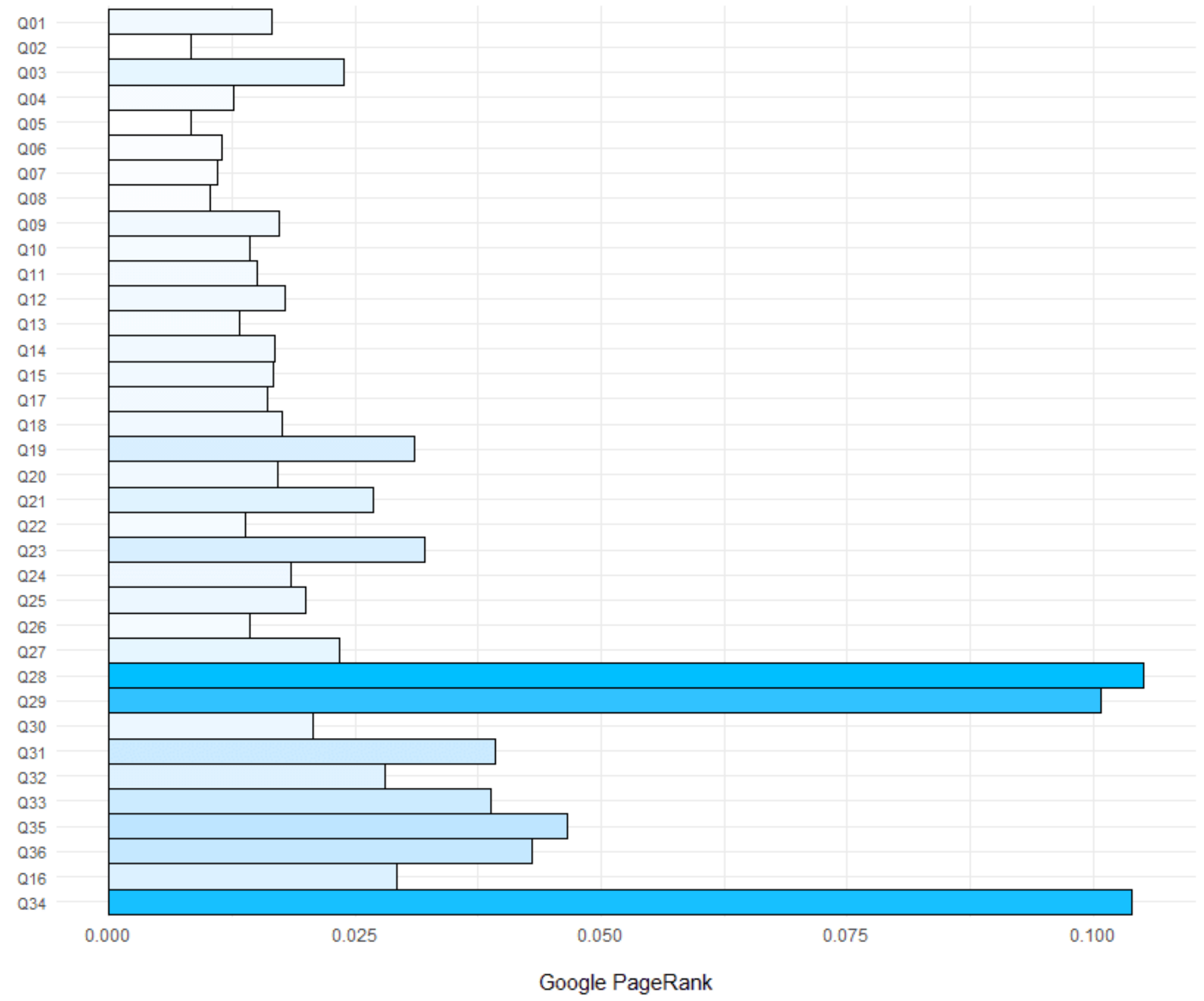

**Figure 6.** Google PageRank centrality of all 36 nodes of the relationship network. The intensity of the blue color of each bar is directly correlated with the centrality value.

## 4.4. Sink to crisis

This section of the result is highlighting the point that everyone should be mindful about in relationships, how it could lead to a critical point or a final node of the system.

According to the C3 subgraph cluster (**Figure 3**), it could be interpreted that when individuals are in the stage of reassurance needed (Q18; *I need a lot of reassurance that I am loved by my partner*), should be cautious that this is the turning point. In the first case, the behavior may convert to positive (will be explained in result point 5). However, in another case, the sense of self-esteem lacking and insecurity might grow stronger. Then, it is either staying in negative behavior or going to the last node (Q34; *When romantic partners disapprove of me, I feel really bad about myself* or Q36; *I resent when my partner is away from me*). Alternatively, in other words, if someone has already passed question 18, there is a great chance that he or she might have to feel insecure till the end of their relationship.

Likewise, from **Figure 3C** (C4), question 20 (Sometimes I feel that I force my partners to show more feeling, more commitment) and question 26 (I find that my partner(s) don't want to get as close as I would like) is a point of awareness. The reason behind it is that those 2 behaviors could convert to a much more overly attached act, namely wanting to merge completely (Q12) or transform to one of the behaviors in C3 which has high chances that will lead to a lack of self-esteem due to the fact that Q18 is excluded. Hence, dropping your expectation may keep your value relationship in a good direction.

On top of that, another point that should be concerned are signs that positive behavior could change to negative. First sign is depending on comfortableness (Q29; I feel comfortable depending on romantic partners.). Second sign is dependence in time of need (Q33; It helps to turn to my romantic partner in times of need.). These 2 signs might result in Q28 and Q32 from cluster 3 respectively. Subsequently, due to the fact that both are not Q18, then the behavior of a person who is not aware of one of these signs may sink to crisis.

## 4.5. Methods to convert negative behaviors to positive behaviors

Due to the fact that each cluster represents a different attachment style, links between C1-C4 to C5 (the only positive behavior cluster in this questionnaire) may be the possible way to make the relationship healthy again. However, there is not every problem that can be fixed and some of those that can be fixed may have less possibility of success than usual (coefficient is less than mean).

First, suppression of maintaining personal space habits (C1; Q01, Q09, Q11, Q21, Q23) could increase feeling comfortable to open up (C5; Q15, Q25, Q27, Q29, Q31, Q33). For instance, repressing Q9, *I don't feel comfortable opening up to romantic partners*, may elevate Q25, *I tell my partner just about everything* (20.57 %). The success rate of this method measured by coefficients of influence or contributions ranges from 9.31 (Q9 → Q27) to 39.35 (Q9 → Q15) %.

The Second class of problems that could be tackled is the anxiety cluster of lacking self-esteem states (C3). There are two major completely positive inter-cluster influences to C3, those from C2 (sum = 1.45430) and C4 (sum = 0.60669). In this case, the only positive behavior that can be changed in this cluster is reassurance required (Q18; *I need a lot of reassurance that I am loved by my partner*). Increasing this act may increase reliance (Q35; *I turn to my partner for many things, including comfort and reassurance.*) which is considered as a positive behavior. Ideally, if individuals, when partners spend time away, are able to start discussing instead of being frustrated and upset, the relationship will get back on the right track. Unfortunately, this path has a small chance to succeed (13.06 percent, less than the median).

Thirdly, minimize expectations (C4; Q26; *I find that my partner(s) don't want to get as close as I would like.*) might strengthen closeness (C5; Q19; *I find it relatively easy to get close with my partner.*) However, the rate of achievement is rather low (9.73 percent).

Lastly, some negative attachment styles may indirectly induce positive attachment styles. Concern about being abandoned (Q02, Q08 and Q14) suppresses the feeling or attitude of not worrying about being abandoned (Q22). Since lack of worry allows an opportunity to maintain personal spaces (Q23; *I prefer not to be too close to romantic partners*), heightened worry of being abandoned urges being closer to partners. As a result, the comfortableness of being close (Q03; *I am very comfortable being close to romantic partners*) will rise. Overall, double negative regulations yield a positive regulation. Therefore, it is of interest to compare these positive effects on Q03 by nodes of C2 with the negative effects by other nodes within C1. By using Formula (2), change in Q03 by Q02 the root node of C2 and change in Q03 by Q05 the root node of C1 can be calculated as sums of products of their respective sequential coefficients in each directed path. Herein, for each root node, 2 paths with the highest coefficient products were selected for demonstration (**Table 4**). The sum of coefficient products of 2 paths from Q02 to Q03 is about 16 folds smaller than those of 2 paths from Q05 to Q03. Even when considering the effects of the more proximal Q22, instead of Q02, on Q03, the product of coefficient is only at 0.0181. While overall influence from Q05 to Q03 is much higher than that from Q02, the influences from Q05 and Q02 are much more comparable and shown in detail in **Table S9**. Since the absolute values of all coefficients are less than 1 (largest: Q05 → Q07 = 0.6277), the indirect or overall influence (coefficient product) of any path from any node to another node will decrease as the number of nodes in the path increases.

| From | To | coeff | | From | To | coeff | | From | To | coeff | | From | To | coeff |
|---|---|---|---|---|---|---|---|---|---|---|---|---|---|---|
| Q02 | Q22 | **-0.4108** | | Q02 | Q08 | **0.5821** | | | | | | | | |
| | | | | Q08 | Q22 | **-0.1976** | | Q05 | Q07 | **0.6277** | | Q05 | Q07 | **0.6277** |
| Q22 | Q23 | 0.1129 | | Q22 | Q23 | 0.1129 | | Q07 | Q23 | 0.1659 | | | | |
| Q23 | Q03 | -0.1603 | | Q23 | Q03 | -0.1603 | | Q23 | Q03 | -0.1603 | | Q07 | Q03 | **-0.2221** |
| **Product** | | 0.0074 | | **Product** | | 0.0021 | | **Product** | | -0.0167 | | **Product** | | -0.1394 |
| | | | | | | | | | | | | | | |
| **Sum** | | 0.0095 | | | | | | **Sum** | | -0.1561 | | | | |

**Table 4**. Two paths from Q02 to Q03 and Q05 to Q03 with the highest coefficient products

## 4.6. Comparison with other computational item-wise studies

In the process of establishing a short form of the ECR, Wei et al. (2008) [21] conducted factor analyses on items of the avoidance and anxiety dimensions separately. From **Table S1**, there are multiple items, especially Q20 and Q28, that cannot be confidently classified by ranking of factors alone. Consequently, we used *k*-means clustering algorithm to partition items of the avoidant and anxious attachment dimensions into 2 (**Figure 7A**) and 3 (**Figure 7B**) clusters respectively. The item lists of the 5 resulting clusters perfectly match those of our 5 subgraphs. The plots by the principal components of the factors are also shown in **Figure S5**.

We also performed *k*-means clustering on the factor data from Lo et al. [22] and Guzmán-González et al. [23]. Most items are grouped with items of the same attachment dimension (**Table 5**). The exceptions manifest only in the clustering of data from Guzmán-González et al. as follows: (i) Q29 is now in an anxiety cluster and (ii) both Q21 and Q22 compose their own outgroup. The cluster plots of the 36 items on the principal components PC1 and PC2 of the original factors from Lo et al. and Guzmán-González et al.

(originally only have 2 factors so the same factors as axis) are shown in **Figure S6A** and **Figure S6B** respectively.

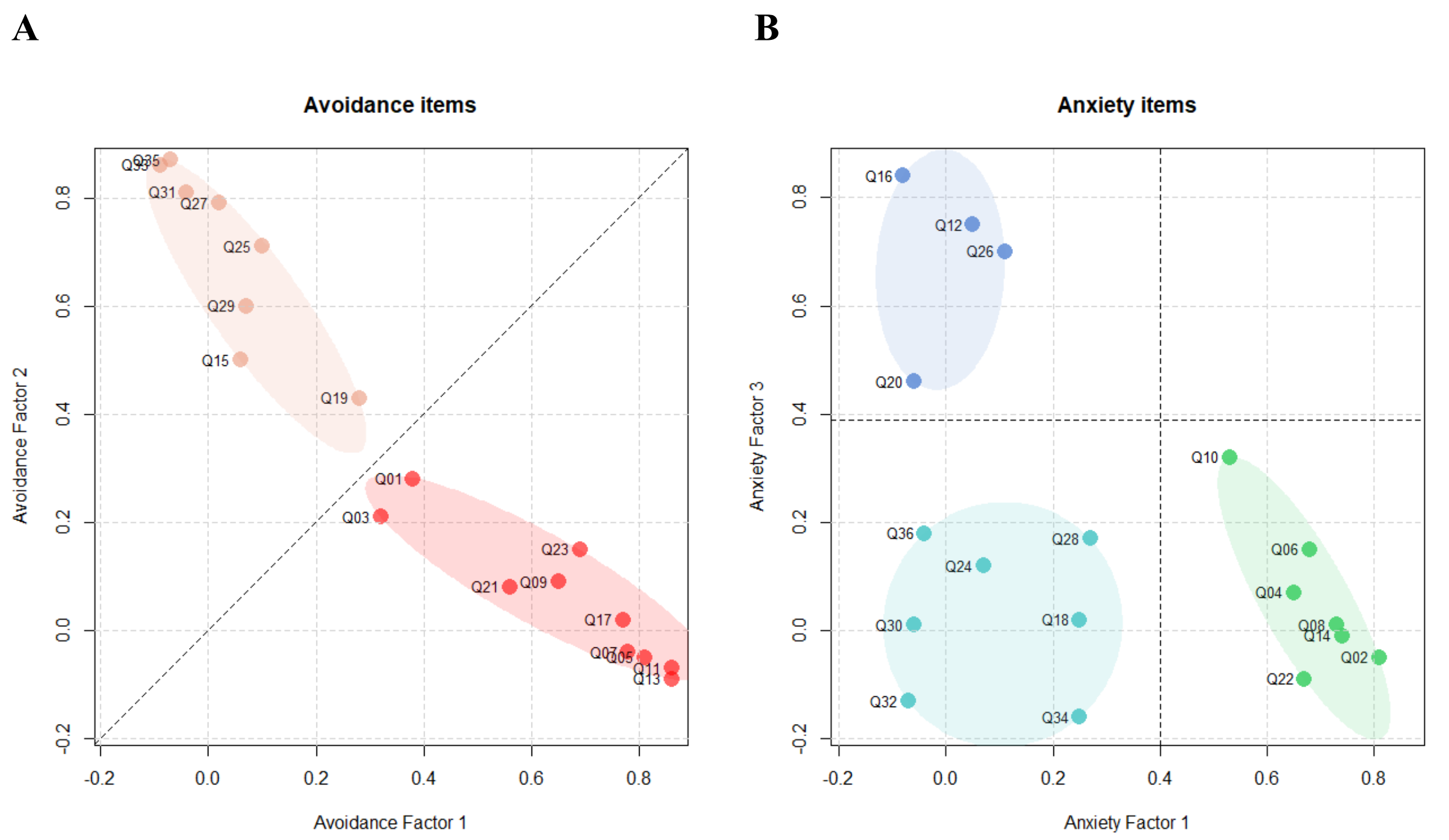

**Figure 7.** Plots of avoidance and anxiety items from Wei et al. (2008)  Avoidance items were plotted by the 2 avoidance factors (A). Anxiety items were plotted by 2 of the 3 anxiety factors (B) that show the most apparent separation between clusters, specifically factor 1 and factor 3 here. Each ellipse is the smallest in area ellipse that can contain all data points of its respective cluster.

| | **Avoidant attachment** | | | | | | | | | | | | | | | | | |
|---|---|---|---|---|---|---|---|---|---|---|---|---|---|---|---|---|---|---|
| [21] | **Q01** | **Q05** | **Q07** | **Q09** | **Q11** | **Q13** | **Q17** | **Q21** | **Q23** | **Q03** | **Q15** | **Q19** | **Q25** | **Q27** | **Q29** | **Q31** | **Q33** | **Q35** |
| [22] | Av 1 | Av 1 | Av 1 | Av 1 | Av 1 | Av 1 | Av 1 | Av 1 | Av 1 | Av 2 | Av 2 | Av 2 | Av 2 | Av 2 | Av 2 | Av 2 | Av 2 | Av 2 |
| [23] | Av 1 | Av 1 | Av 1 | Av 1 | Av 1 | Av 1 | Av 1 | Out | Av 1 | Av 1 | Av 2 | Av 2 | Av 2 | Av 2 | Ax 1 | Av 2 | Av 2 | Av 2 |
| | **Anxious attachment** | | | | | | | | | | | | | | | | | |
| [21] | **Q02** | **Q04** | **Q06** | **Q08** | **Q10** | **Q14** | **Q22** | **Q18** | **Q24** | **Q28** | **Q30** | **Q32** | **Q34** | **Q36** | **Q12** | **Q16** | **Q20** | **Q26** |
| [22] | Ax 1 | Ax 1 | Ax 1 | Ax 1 | Ax 3 | Ax 1 | Ax 1 | Ax 3 | Ax 2 | Ax 3 | Ax 2 | Ax 2 | Ax 3 | Ax 2 | Ax 3 | Ax 3 | Ax 3 | Ax 3 |
| [23] | Ax 1 | Ax 1 | Ax 1 | Ax 1 | Ax 1 | Ax 1 | Out | Ax 1 | Ax 2 | Ax 2 | Ax 1 | Ax 1 | Ax 1 | Ax 1 | Ax 2 | Ax 2 | Ax 2 | Ax 2 |

**Table 5.** The *k*-means clustering results of data from the 3 factor analyses studies. Item clusters of Wei et al. [21], which completely match our subgraph partitions, are compared to the item clusters of Lo et al. [22] and Guzmán-González et al. [23]. Underlined items are the outliers (lying outside the 95% confidence ellipse) of their respective groups. The outgroup (Out) from Guzmán-González et al. only has 2 items: Q21 and Q22.

A study by McWilliams and Fried [24] described in the related work section uses network analysis to model relationships between attachment items. We can compare the results of our Bayesian network analysis to their model by finding correlations between our estimated parameters and their correlation coefficients from the weighted adjacency matrix of the romantic partner cohort (their Study 2). In this case, we exclude any node pairs with the RSQ item 10 from the correlation since it cannot be traced back to any items in the original ECR. For all shared node pairs excluding those lacking edges in both studies (N = 26, df = 24), the Pearson correlation coefficient is at 0.823 and the *p*-value is significant at $2.48 \times 10^{-7}$ (**Table S10C**). Even for only the pairs with defined edges in both studies (N = 12, df = 10), the correlation coefficient is at 0.626 and the *p*-value is still statistically significant at 0.0294 (**Table S10D**).

# 5. DISCUSSION AND CONCLUSION

Causal relationships between 36 questions in the questionnaire *Experiences in Close Relationships Scale, a test of attachment style,* were modeled from the online response data as Bayesian network. The dependency between each node and its parents was modeled as a linear function. We have shown that the baseline values or intercepts of positive or secure attachment items are higher than those of negative attachment. This agrees with previous studies where the majority of relationships are in a secure attachment state [31].

The 36 items are grouped into 5 clusters, 2 of which are avoidance attachment and the other 3 are anxiety attachment. This is congruent with previous reports and reviews that attachment dimensions are relatively independent [9][21]. Nevertheless, connecting arcs between clusters especially of items of different attachment dimensions also confirm the notion that attachment styles may change over time given a sufficient stimulus or specific environment. The highest coefficient sum of between-cluster arcs is from C1 to C5. There is no arc between C1 and C3 or C4, and between C2 and C5.

The root nodes of the system are Q05 from the avoidance C1 and Q02 from the anxiety C2. This confirms the relative independence between the avoidance and anxiety dimensions. Interestingly, they also have the two highest baselines among all negative items. Q16, Q34, and Q36 are the terminal nodes or final outcomes of the system with Q34 having the second highest PageRank score among all items. All 3 of them are of the anxiety attachment dimension and there is no terminal node of the avoidance dimension. However, these 2 root nodes and 3 terminal nodes are not unanimously selected by the 3 factor analysis studies as one of item subsets or short-forms to represent the original 36-item-ECR. Wei et al. [21] identified Q16 (**Table S1**). Lo et al. [22] has Q02 and Q36 (**Table S2**). Guzmán-González et al. [23] has a list including Q05 and Q20 (**Table S3**). Only Q17 (C1) and Q27 (C5) of the avoidance dimension and Q6 (C2) and Q18 (C3) of the anxiety dimension are shared in all 3 studies. In terms of node importance by connection, Lo et al. [22] and Guzmán-González et al. [23] have both Q09 and Q23, the two nodes with the highest betweenness centrality, in their lists of representative items.

Fascinatingly, classification of items by *k*-means clustering on the factors identified by Wei et al. (**Table S1**), specifically 2 avoidance and 3 anxiety factors, yields a complete match with our subgraphs divided by a *cluster_walktrap* algorithm. Factor scores and relative distance may also indicate multiple characteristics of items that can be explained by our Bayesian network model. For instance, Q28 belongs to C3 but has the highest value in C2-associated anxiety factor 1 (**Figure 7B**) or least distance from C2 (**Figure S5B**). Our network model shows that Q28 is heavily influenced by Q14 from C2 with a coefficient of 0.383 (**Table S8**), which is even higher than an influence coefficient of 0.199 from Q30 of the same cluster C3 (**Table S7**). Similarly, a relatively high anxiety factor 3 of the item Q10 (**Figure 7B**) can be explained by a high sum of influence coefficients from Q10 to the C4.

Factor analysis on the avoidant and anxious attachment dimensions separately has an advantage of effective item clustering within each of the 2 subscales, but it cannot reveal potential connections between the 2 attachment dimensions. Work by Lo et al. [22] (**Table S2**) and Guzmán-González et al. [23] (**Table S3**), which performed factor analysis on all 36 items together, both identified 3 anxious attachment items Q12, Q16, and Q26 to have prominent avoidant factors. This is congruent with our finding that these 3 items together with Q20 belong to the cluster C4 that represents the state of being overly close, the opposite of avoidance. An important example from Lo et al. is Q22 having a negative factor 2, which is an avoidance factor, of -0.23. Our study showed that a positive anxiety item Q22 positively influences Q23. Lo et al., unlike our study, reversed the key of all positive items prior to the factor analysis. Therefore, it naturally follows that a reversed key Q22 would exhibit a negative value of factor 2 which is the highest factor of Q23. In a strikingly similar manner, Q29 having a negative factor 1 of -24 is consistent with our BN model that Q29 positively influences Q28, a factor-1-associated item (**Table S8**). While the 2-factor analysis by Guzmán-González et al. cannot capture any distinctive features of Q22, it clearly demonstrates that Q29 has a predominant anxiety factor. This is consistent with an influence of Q29 on Q28 revealed by our BN model and its exorbitant high PR score together with Q28 and Q34. Surprisingly, factor analyses by neither of the 2 studies do not show any evidence of a positive inter-cluster influence from Q18 (C2) to Q35 (C5) (**Table S8**).

Correlation between the influence weights of this study and the partial correlation weights of one network study [24] for a set of shared edges demonstrated similar results between two different network analysis approaches and partly validated our study. Items 2 and 9 of the RSQ used in the network study do not share the exact statements with any of the 36 ECR items and consequently were assigned to the ECR items with the most similar statements Q25 and Q18 respectively (**Table S10B**). As expected of the partial but not exact similarities with their respective counterparts, excluding both (2:Q25) and (9:Q18) from the analysis increases the correlation coefficient to 0.752 but the *p*-value is only 0.085 due to the small sample size (df = 4) (data not shown).

This is the first research that uses Bayesian network analysis to analyse items from the ECR -- *Experiences in Close Relationships Scale, a test of attachment style*. We uncover multiple key findings as follows.

1. Baselines show that in the first stage individuals tend to have positive behavior.
2. Questions in the survey are divided into 5 clusters which are theoretically correct due to attachment style theory.

3. There are transactions between clusters in different possibilities and the main path is C1 to C5.
4. Q16, Q34 and Q36 are the last nodes of the system. Therefore, Q18,Q20 and Q26 are points of awareness.
5. An unhealthy relationship behavior could be converted to a healthy relationship manner by intervention of relevant attachment states represented by ECR items.
6. Bayesian network result is consistent with previous factor analysis studies and demonstrates that causal relationships can help explain the factor profiles of many attachment items .

## ACKNOWLEDGEMENTS

This research could not be possible without: (i) Guidance from Dr. Kobchai Duangrattanalert who did most of the coding to the fullest extent and always pushed me to be the best version I could be; (ii) Genuine feedback, hard work on editing, and inspiring brainstorming from Dr. Chanati Jantrachotechatchawan; (iii) Advice related to programming and machine learning from Pongpak Manoret; (iv) Funding from DPST (Development and Promotion of Science and Technology Talents Project); (v) Support from Triam Udom Suksa School; (vi) Opportunity from IPST (The Institute for the Promotion of Teaching Science and Technology); and (vii) As well as the reason why the study started in the first place: love: a complex emotion that gives the best motivation for me to complete this project and for us to venture in this complicated world.

## REFERENCES

1. Kposowa, A.J. (2000) Marital status and suicide in the National Longitudinal Mortality Study. J. Epidemiol. Community. Health. Apr;54(4), 254-61. https://dx.doi.org/10.1136/jech.54.4.254. PMID: 10827907; PMCID: PMC1731658.
2. Department of Economic and Social Affairs, United Nations. (2020). Demographic Yearbook 70th Issue - 2019. New York: United Nations Publication; NUPTIALITY: Table 23 - Marriages and crude marriage rates, by urban/rural residence: 2015 - 2019. Available from: https://unstats.un.org/unsd/demographic/products/dyb/dyb2011/Table23.pdf. https://unstats.un.org/unsd/demographic-social/products/dyb/dyb_2019/
3. Department of Economic and Social Affairs, United Nations. (2020). Demographic Yearbook 70th Issue - 2019. New York: United Nations Publication; DIVORCE: Table 25 - Divorces and crude divorce rates by urban/rural residence: 2015 - 2019. [cited May 2021]. Available from: https://unstats.un.org/unsd/demographic/products/dyb/dyb2011/Table25.pdf. https://unstats.un.org/unsd/demographic-social/products/dyb/dyb_2019/
4. Statistics Korea (KOSTAT). (2021). Statistics Korea; 2010. Marriage and Divorce; Available from: http://kostat.go.kr/portal/eng/pressReleases/8/11/index.board.
5. National Center for Health Statistics. (2021) Centers for Disease Control and Prevention; Marriage and Divorce. Available from: https://www.cdc.gov/nchs/fastats/marriage-divorce.htm.
6. Australian Bureau of Statistics. (2021). Australian Bureau of Statistics. Marriages and Divorces, Australia. Available from:

https://www.abs.gov.au/statistics/people/people-and-communities/marriages-and-divorces-australia/latest-release.

7. Pei-Ju, T. (2019) An average of 149 couples file for divorce per day in Taiwan in 2018: data. Taiwan News. Available from: https://www.taiwannews.com.tw/en/news/3730261.
8. Taormina, R.J., Gao, J.H. (2013) Maslow and the motivation hierarchy: measuring satisfaction of the needs. Am. J. Psychol. 126(2), 155-77. http://doi.org/10.5406/amerjpsyc.126.2.0155.
9. Brennan, K.A., Clark, C.L., Shaver, P.R. (1998) Self-report measurement of adult attachment: An integrative overview. In Simpson JA & Rholes WS (Eds.), Attachment theory and close relationships. pp. 46–76. The Guilford Press.
10. Slatcher, R.B., Selcuk, E. (2017). A Social Psychological Perspective on the Links between Close Relationships and Health. Curr. Dir. Psychol. Sci. Feb;26(1), 16-21. doi: 10.1177/0963721416667444. Epub 2017 Feb 8. PMID: 28367003; PMCID: PMC5373007.
11. Feeney BC. The dependency paradox in close relationships: accepting dependence promotes independence. J Pers Soc Psychol. 2007 Feb;92(2), 268-85. doi: 10.1037/0022-3514.92.2.268. PMID: 17279849.
12. Weidman, A.C., Sowden, W.J., Berg, M.K., and Kross, E. (2020). Punish or Protect? How Close Relationships Shape Responses to Moral Violations. Pers. Soc. Psychol. Bull. May;46(5), 693-708. doi: 10.1177/0146167219873485. Epub 2019 Sep 19. PMID: 31535954.
13. Hofmann, W., Finkel, E.J., and Fitzsimons, G.M. (2015). Close relationships and self-regulation: How relationship satisfaction facilitates momentary goal pursuit. J. Pers. Soc. Psychol. Sep;109(3):434-52. doi: 10.1037/pspi0000020. Epub 2015 Jun 29. Erratum in: J Pers Soc Psychol. 2016 Apr;110(4), 591. PMID: 26121524.
14. Ishii K. Conflict management in online relationships. Cyberpsychol Behav Soc Netw. 2010 Aug;13(4), 365-70. doi: 10.1089/cyber.2009.0272. PMID: 20712494.
15. Birditt, K.S., Sherman, C.W., Polenick, C.A., Becker, L., Webster, N.J., Ajrouch, K.J., and Antonucci, T.C. (2020). So Close and Yet So Irritating: Negative Relations and Implications for Well-being by Age and Closeness. J. Gerontol. B. Psychol. Sci. Soc. Sci. Jan 14;75(2), 327-337. doi: 10.1093/geronb/gby038. PMID: 29596623; PMCID: PMC7179808.
16. Strelan, P., Karremans, J.C., and Krieg, J. (2017). What determines forgiveness in close relationships? The role of post-transgression trust. Br. J. Soc. Psychol. Mar;56(1), 161-180. doi: 10.1111/bjso.12173. Epub 2016 Nov 16. PMID: 27862012.
17. Oon-Arom, A., Wongpakaran, T., Satthapisit, S., Saisavoey, N., Kuntawong, P., and Wongpakaran, N. (2019). Suicidality in the elderly: Role of adult attachment. Asian J. Psychiatr. Aug;44, 8-12. doi: 10.1016/j.ajp.2019.07.014. Epub 2019 Jul 5. PMID: 31302442.
18. González-Ortega, E., Orgaz, B., Vicario-Molina, I., and Fuertes, A. (2021). Adult Attachment, Conflict Resolution Style and Relationship Quality among Spanish Young-adult Couples. Span J Psychol. 2021 Feb 5;24:e5. doi: 10.1017/SJP.2021.4. PMID: 33541453.
19. Bowlby, J. (1980). Attachment and Loss: Loss, Sadness and Depression, Vol. 3, Basic, New York.
20. Ainsworth, M.D.S., Blehar, M.C., Waters, E., and Wall, S.N. (1978) Patterns of Attachment: Observations in the Strange Situation and at Home. Hillsdale, N: Erlbaum.
21. Wei, M., Russell, D.W., Mallinckrodt, B., and Vogel, D.L. (2007). The Experiences in Close Relationship Scale (ECR)-short form: reliability, validity, and factor structure. J. Pers. Assess. 2007 Apr;88(2), 187-204. doi: 10.1080/00223890701268041. PMID: 17437384.
22. Lo, C., Walsh, A., Mikulincer, M., Gagliese, L., Zimmermann, C., and Rodin, G. (2009). Measuring attachment security in patients with advanced cancer: psychometric properties of a modified and brief Experiences in Close Relationships scale. Psychooncology May;18(5), 490-9. doi: 10.1002/pon.1417. PMID: 18821528.

23. Guzmán-González, M., Rivera-Ottenberger, D., Brassard, A., Spencer, R., and Lafontaine, M.F. (2020). Measuring adult romantic attachment: psychometric properties of the brief Spanish version of the experiences in close relationships. Psicol. Reflex. Crit. Jun 5;33(1), 9. doi: 10.1186/s41155-020-00145-w. PMID: 32542456; PMCID: PMC7295914.
24. McWilliams, L. and Fried, E. (2019). Reconceptualizing adult attachment relationships: A network perspective. Personal Relationships 26. doi: 10.1111/pere.12263.
25. Fraley, R.C., Heffernan, M.E., and Vicary, A.M., Brumbaugh, C.C. (2011). The Experiences in Close Relationships-Relationship Structures questionnaire: a method for assessing attachment orientations across relationships. Psychol. Assess. Sep;23(3), 615-25. doi: 10.1037/a0022898. PMID: 21443364.
26. Kim, J.H. and Pearl, J. (1983). A computational model for causal and diagnostic reasoning in inference engines. In Proceeding of the 8th International Joint Conference on Artificial Intelligence. Karlsruhe, West Germany, 190-193.
27. Silvia, S. and Ron, K. (2007). Bayesian networks of customer satisfaction survey data. Department of Economics University of Milan Italy, Departmental Working Papers. 36. 10.1080/02664760802587982.
28. Chakraborty, S., Mengersen, K., Fidge, C. et al. (2016). A Bayesian Network-based customer satisfaction model: a tool for management decisions in railway transport. Decis. Anal. 3, 4. https://doi.org/10.1186/s40165-016-0021-2
29. Golia, S., Pedrazza, M., Berlanda, S., and Trifiletti, E. (2017). The interplay of attachment avoidance and anxiety in affecting nurse's caregiving style and emotional exhaustion. TPM. 24, 437-457. doi: 10.4473/TPM24.3.9.
30. Nagarajan, R., Lèbre, S., & Scutari, M. (2013). Bayesian networks in R: With applications in systems biology. New York: Springer.
31. Mikulincer, M. and Shaver, P. R. (2003). The attachment behavioral system in adulthood: activation, psychodynamics, and interpersonal processes. Adv. Exp. Soc. Psychol. 35, 53–152. doi: 10.1016/S0065-2601(03)01002-5.

# SUPPLEMENTARY MATERIALS

## Supplementary method: Demonstration of indirect influence calculation

To demonstrate the calculation of indirect influences, consider the following directed graph example of 5 nodes with a root node $p$ and an end node $t$ (**Figure S1**). For sake of simplicity, in this example, we will also use $p, q, r, s, t$ to represent both nodes and variables instead of the more standard notations $v_p, v_q, v_r, v_s, v_t$ and $x_p, x_q, x_r, x_s, x_t$ respectively.

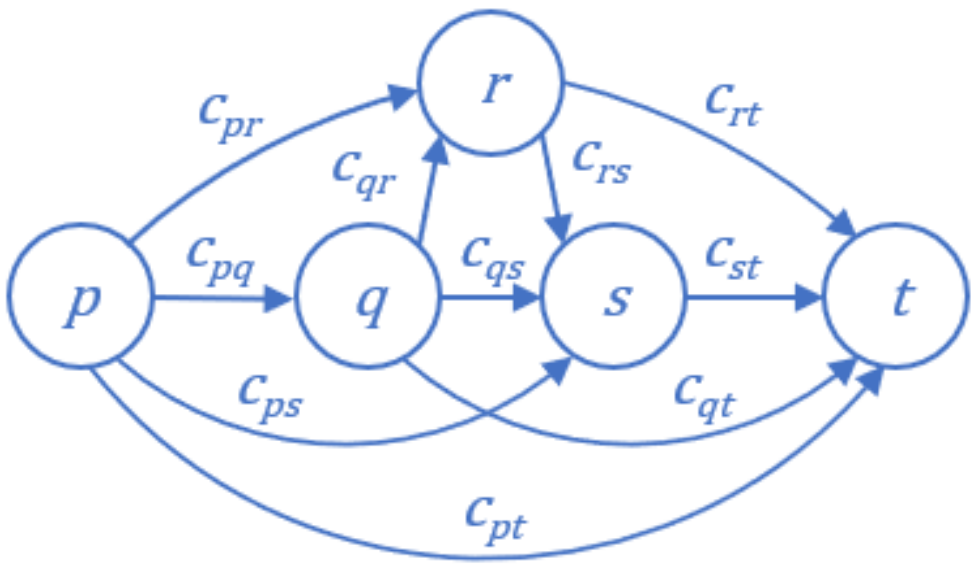

**Figure S1.** An example of a directed acyclic graph of Bayesian network with estimated coefficients as weights of arcs

If we want to find total influence from $q$ to $t$. First calculate the derivative or sum of coefficient products corresponding to a set of paths from each parent node of $t$, followed by those of a set of paths from a parent of each of the last considered parents, until any parent nodes of $q$ are reached.

$$\frac{dt}{dq} = c_{rt}\frac{dr}{dq} + c_{st}\frac{ds}{dq} + c_{qt}\frac{dq}{dq} + c_{pt}\frac{dp}{dq} + \frac{d}{dq}b_t \quad = c_{rt}\frac{dr}{dq} + c_{st}\frac{ds}{dq} + c_{qt} + 0 + 0 \;;$$

$$\frac{dr}{dq} = c_{qr}\frac{dq}{dq} + c_{pr}\frac{dp}{dq} + \frac{d}{dq}b_r \quad = c_{qr} + 0 + 0 \;;$$

$$\frac{ds}{dq} = c_{qs}\frac{dq}{dq} + c_{rs}\frac{dr}{dq} + c_{ps}\frac{ds}{dp} + \frac{d}{dq}b_s \quad = c_{qs} + c_{rs}\frac{dr}{dq} + 0 + 0 \;;$$

$$\frac{dt}{dq} = c_{rt}\,(c_{qr}) + c_{st}\,(c_{qs} + c_{rs}\,(c_{qr})) + c_{qt}$$

$$= (c_{rt}\,c_{qr}) + (c_{rs}\,c_{qs}) + (c_{st}\,c_{rs}\,c_{qr}) + (c_{qt}) \qquad = (c_{qr}\,c_{rt}) + (c_{qs}\,c_{st}) + (c_{qr}\,c_{rs}\,c_{st}) + (c_{qt})$$

Each of these 4 terms corresponds to the coefficient product of one of the following 4 directed paths from $q$ to $t$: $\{(e_{qr}, e_{rt}),\ (e_{qs}, e_{st}),\ (e_{qr}, e_{rs}, e_{st}),\ (e_{qt})\}$

| **Avoidance** | | **Anxiety** | | |
|---|---|---|---|---|
| **Factor 1** | **Factor 2** | **Factor 1** | **Factor 2** | **Factor 3** |
| Q11 [*r* = .68] (**.86**, -.07) | Q35 [*r* = .65] (-.07, **.87**) | Q02 [*r* = .68] (**.81**, .01, -.05) | Q32 [*r* = .52] (-.07, **.87**, -.13) | Q16 [*r* = .56] (-.08, .01, **.84**) |
| Q13 [*r* = .66] (**.86**, -.09) | Q33 [*r* = .65] (-.09, **.86**) | Q14 [*r* = .69] (**.74**, .05, -.01) | Q30 [*r* = .59] (-.06, **.81**, .01) | Q12 [*r* = .55] (.05, -.07, **.75**) |
| Q05 [*r* = .66] (**.81**, -.05) | Q31 [*r* = .64] (-.04, **.81**) | Q08 [*r* = .72] (**.73**, .02, .01) | Q36 [*r* = .52] (-.04, **.54**, .18) | Q26 [*r* = .61] (.11, -.03, **.70**) |
| Q07 [*r* = .65] (**.78**, -.04) | Q27 [*r* = .67] (.02, **.79**) | Q06 [*r* = .69] (**.68**, -.03, .15) | Q34 [*r* = .51] (.25, **.52**, -.16) | Q20 [*r* = .59] (-.06, .35, **.46**) |
| Q17 [*r* = .65] (**.77**, .02) | Q25 [*r* = .68] (.10, **.71**) | Q22 [*r* = .56] (**.67**, -.05, -.09) | Q18 [*r* = .66] (.25, **.52**, .02) | |
| Q23 [*r* = .65] (**.69**, .15) | Q29 [*r* = .57] (.07, **.60**) | Q04 [*r* = .65] (**.65**, .01, .07) | Q24 [*r* = .55] (.07, **.51**, .12) | |
| Q09 [*r* = .64] (**.65**, .09) | Q15 [*r* = .49] (.06, **.50**) | Q10 [*r* = .66] (**.53**, -.07, .32) | | |
| Q21 [*r* = .65] (**.56**, .08) | Q19 [*r* = .61] (.28, **.43**) | Q28 [*r* = .57] (**.27**, .24, .17) | | |
| Q01 [*r* = .57] (**.38**, .28) | | | | |
| Q03 [*r* = .47] (**.32**, .21) | | | | |

**Table S1.** Factors of 36 ECR items, modified from Wei et al. [21]. Classification of each ECR item by its factor with the highest absolute value. In each factor column, the items are decreasingly ordered by their corresponding factor values. Light yellow, orange, and violet highlight the items of which the second-highest factor is higher than 50%, 75%, and 87.5% of the highest factor respectively. Items with names in red are those selected for the short-form by the study.

| Anxiety | Avoidance | Avoidance | Anxiety |
|---|---|---|---|
| **Factor 1** | **Factor 2** | **Factor 3** | **Factor 4** |
| Q30 (**.84**, -.18, .09, -.01) | Q07 (-.08, **.76**, .01, .01) | Q31 (-.11, .02, **.66**, .022) | Q02 (.06, .02, .01, **.68**) |
| Q32 (**.83**, -.13, .05, -.10) | Q13 (.17, **.75**, .05, .12) | Q27 (.16, -.04, **.65**, .22) | Q08 (.09, .11, -.03, **.59**) |
| Q36 (**.68**, .02, -.03, .03) | Q17 (.14, **.71**, .16, -.14) | Q33 (.07, .03, **.62**, .07) | Q04 (.28, -.01, .03, **.52**) |
| Q24 (**.58**, -.01, .02, .04) | Q23 (.14, **.64**, .18, .18) | Q35 (-.07, .04, **.61**, .00) | Q06 (.21, .32, -.06, **.43**) |
| Q28 (**.53**, .09, -.06, .11) | Q09 (.00, **.63**, .18, -.05) | Q25 (-.02, .01, **.59**, -.05) | Q14 (.43, -.15, .12, **.51**) |
| Q20 (**.50**, .09, -.05, .12) | Q05 (-.12, **.63**, .06, .26) | Q15 (-.12, .13, **.51**, .05) | Q22 (.18, -.23, .14, **.37**) |
| Q34 (**.46**, .13, -.02, .02) | Q11 (-.17, **.61**, -.04, .10) | Q19 (.03, .15, **.47**, -.08) | |
| Q18 (**.45**, .00, -.02, .31) | Q01 (.12, **.43**, 20, .06) | Q03 (.10, .21, **.40**, -.07) | |
| Q12 (**.45**, .24, -.19, .13) | Q21 (-.04, **.21**, .05, .07) | Q29 (-.24, .06, **.36**, .14) | |
| Q26 (**.43**, .31, -.02, .09) | | | |
| Q16 (**.40**, .25, -.11, .14) | | | |
| Q10 (**.35**, .11, .01, .28) | | | |

**Table S2.** Factors of 36 ECR items, modified from Lo et al. [22]. Classification of each ECR item by a factor with the highest absolute value. In each factor column, the items are decreasingly ordered by their corresponding factor values. Light yellow and orange highlight the items of which the second-highest factor is higher than 50% and 75% of the highest factor respectively. Items with names in red are those selected for the short-form by the study.

| Avoidance | | Anxiety | |
|---|---|---|---|
| Q23 [*r* = .61] (.15, **.66**) | Q33 [*r* = .49] (.24, **-.54**) | Q30 [*r* = .64] (**.68**, .08) | Q14 [*r* = .56] (**.56**, -.00) |
| Q05 [*r* = .60] (.20, **.63**) | Q07 [*r* = .50] (.23, **.53**) | Q32 [*r* = .58] (**.63**, -.02) | Q10 [*r* = .50] (**.55**, -.10) |
| Q09 [*r* = .57] (.20, **.61**) | Q19 [*r* = .48] (.10, **-.53**) | Q18 [*r* = .56] (**.63**, -.12) | Q04 [*r* = .50] (**.54**, -.13) |
| Q17 [*r* = .55] (.19, **.61**) | Q31 [*r* = .47] (.17, **-.52**) | Q24 [*r* = .58] (**.61**, .13) | Q12 [*r* = .52] (**.52**, .26) |
| Q13 [*r* = .56] (.28, **.60**) | Q01 [*r* = .48] (.13, **.51**) | Q06 [*r* = .58] (**.60**, .05) | Q34 [*r* = .46] (**.50**, .04) |
| Q27 [*r* = .55] (.11, **-.59**) | Q15 [*r* = .45] (.16, **-.50**) | Q20 [*r* = .57] (**.60**, .15) | Q28 [*r* = .50] (**.50**, .14) |
| Q35 [*r* = .53] (.29, **-.59**) | Q25 [*r* = .46] (.22, **-.49**) | Q36 [*r* = .54] (**.60**, -.10) | Q26 [*r* = .50] (**.47**, .35) |
| Q11 [*r* = .54] (.32, **.56**) | Q29 [*r* = .11] (**.33**, -.14) | Q08 [*r* = .58] (**.60**, -.10) | Q16 [*r* = .39] (**.39**, .18) |
| Q03 [*r* = .49] (-.13, **.55**) | Q21 [*r* = .15] (.10, **.13**) | Q02 [*r* = .55] (**.56**, -.05) | Q22 [*r* = .18] (**-.16**, .05) |

**Table S3.** Factors of 36 ECR items, modified from Guzmán-González et al. [23]. For each item, correlation with the total score (in square bracket) and values of the anxiety (factor 1) and avoidance (factor 2) factors are listed. A factor with the highest absolute value is bolded. In each factor column, the items are decreasingly ordered by their attachment dimension-corresponding factor values. Light yellow highlights the items of which the second-highest factor is higher than 50% of the highest factor. Red highlights the items of which dimension-corresponding factors have absolute values less than 0.20. Items with names in red are those selected for the short-form by the study.

**A**

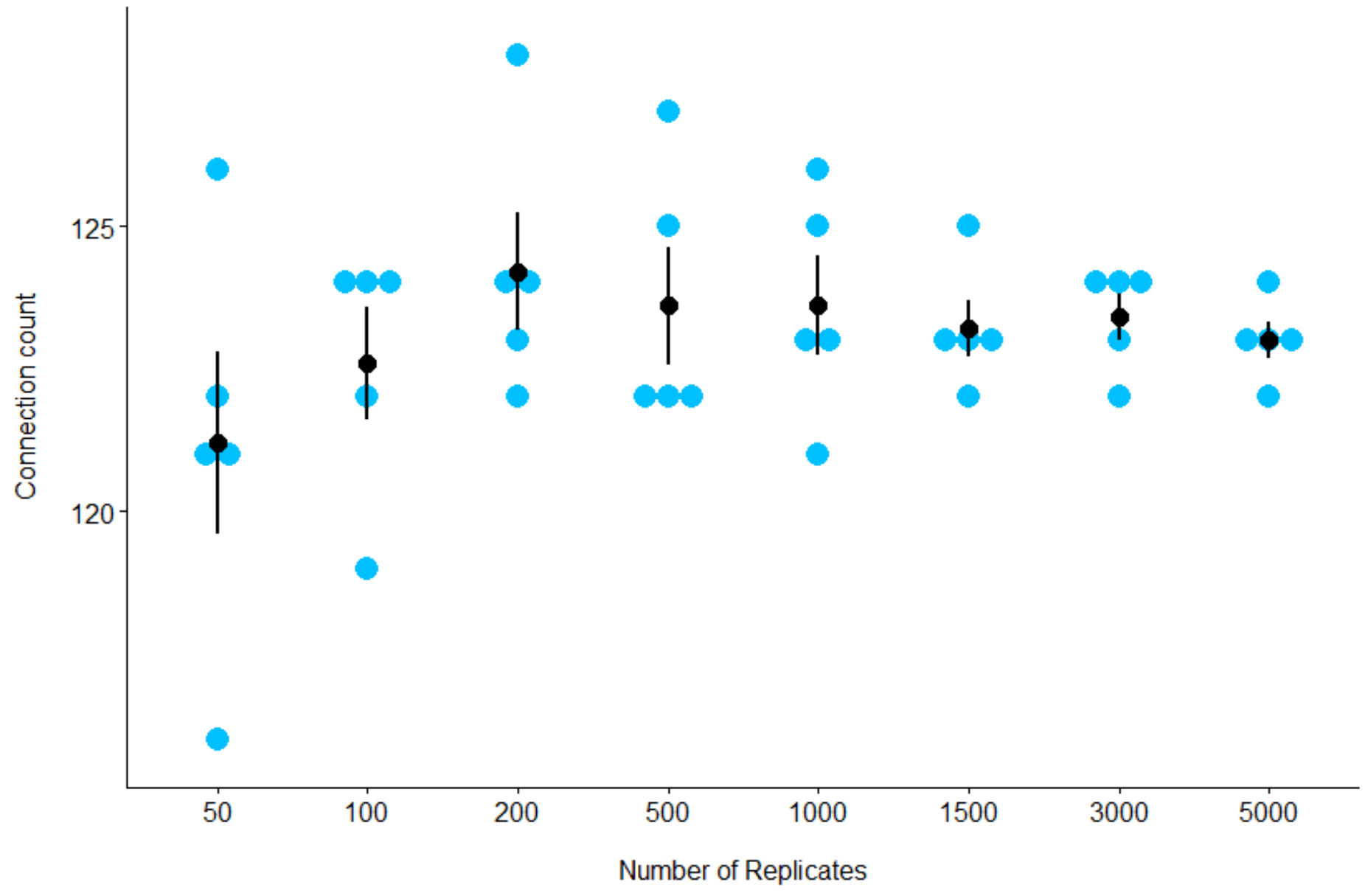

**B**

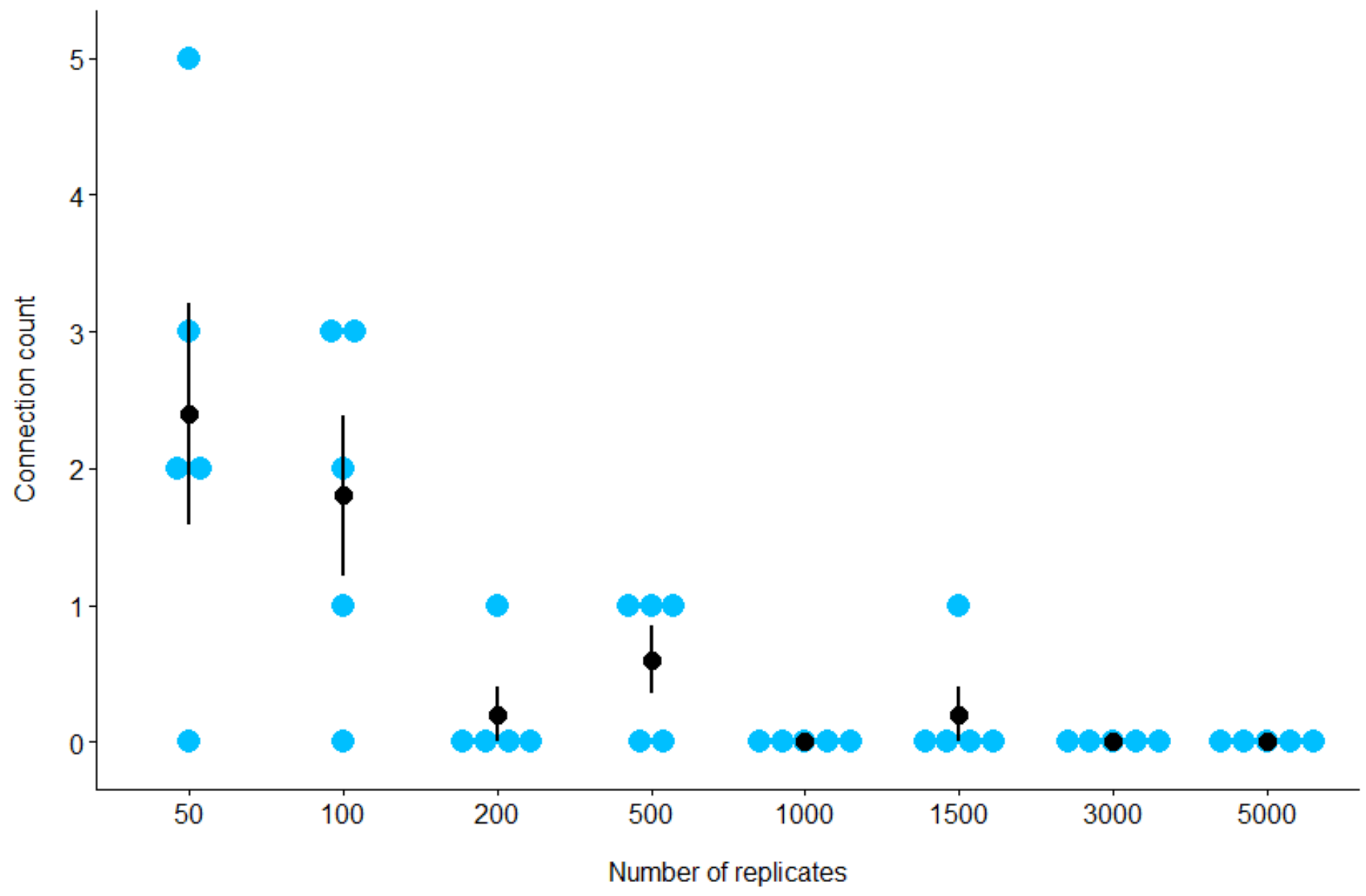

**Figure S2.** Counts of directed and undirected graph connections after training. Counts of directed graph connections (A) and counts of undirected graph connections (B) by the number of bootstrap replicates. B) reveal that training for 3000 epochs is sufficient to achieve stable counts of 123 directed connections and 0 undirected connections. Data are represented as mean ± standard deviation (n=5).

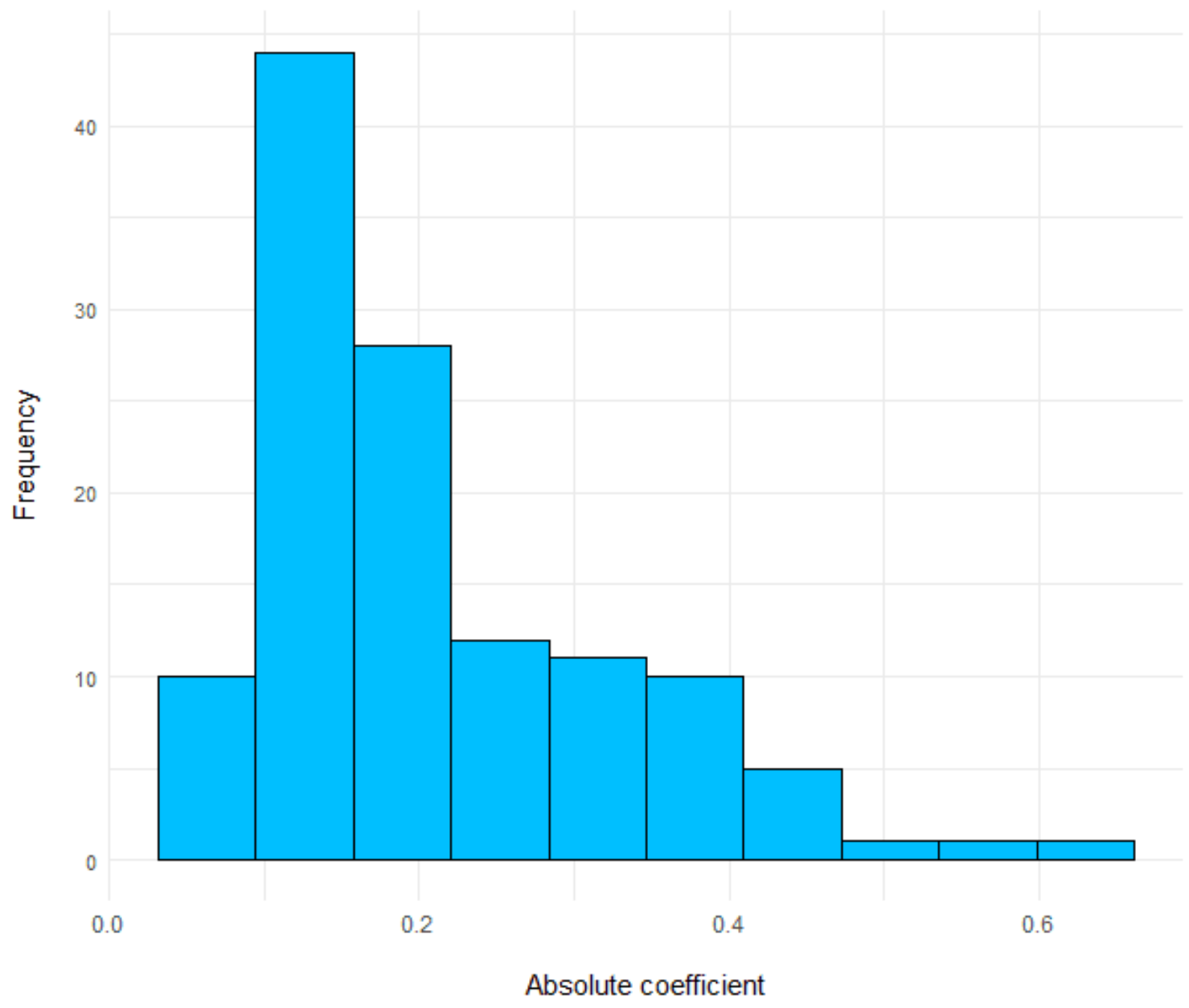

**Figure S3.** Frequency of absolute coefficients of the relationship network. The median is at 0.17217.

**A**

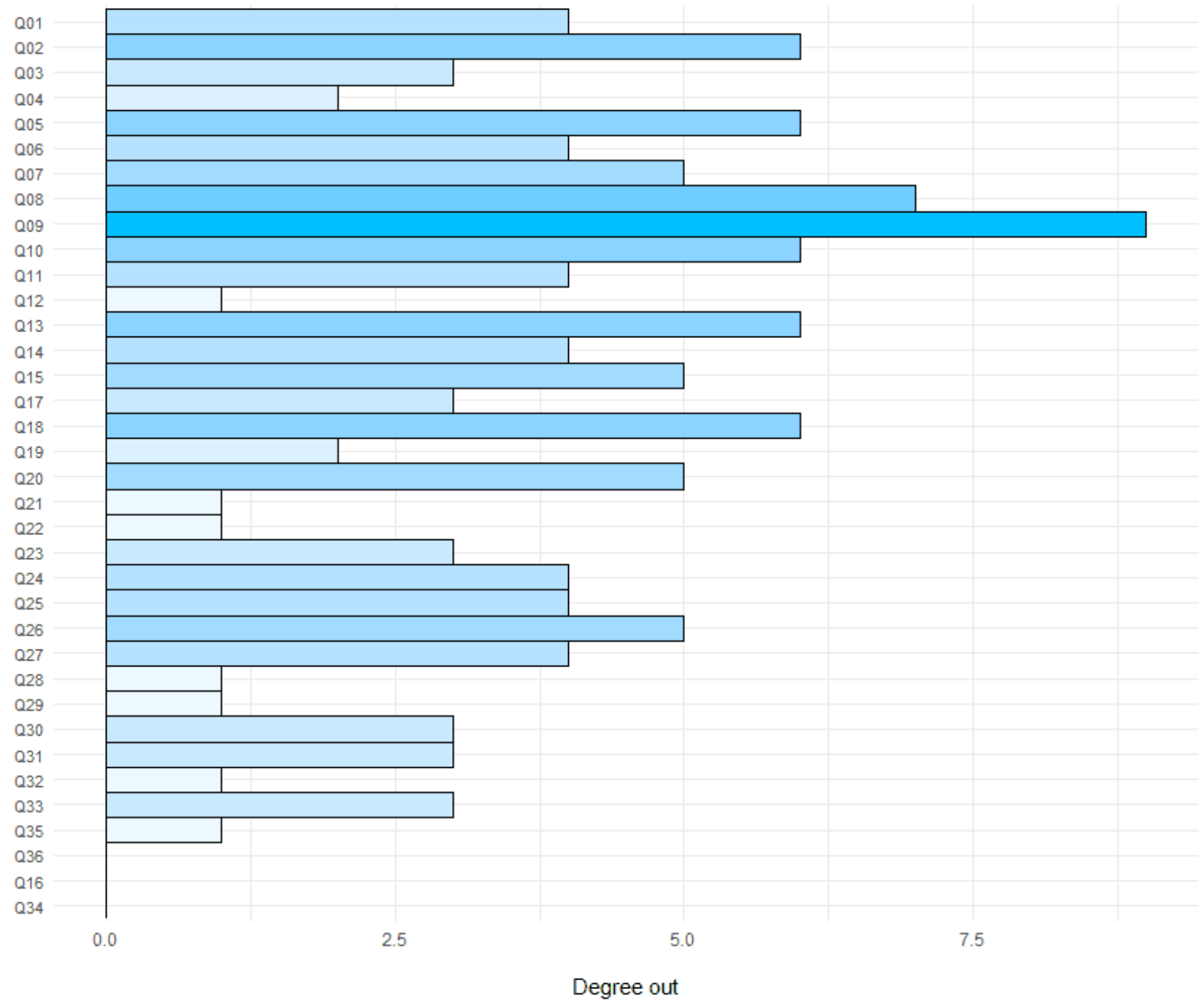

**B**

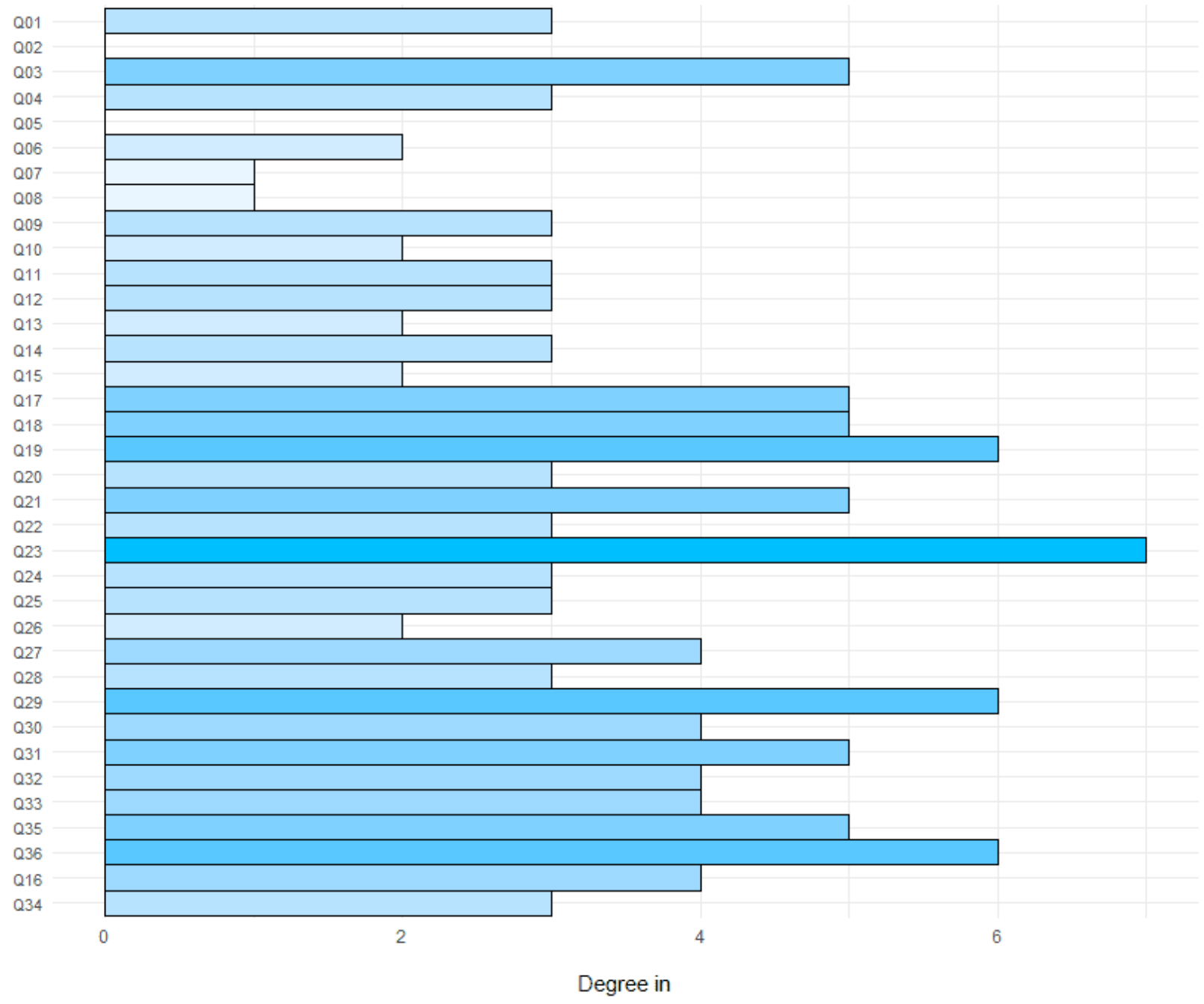

**Figure S4.** Degree out (A) and degree in (B) centrality of all 36 nodes of the network. The intensity of the blue color corresponds with the degree value. The maximum degree out is from node Q09 and the maximum degree in is to node Q23.

| Q # | Intercept | Stdv | Questions / Items |
|---|---|---|---|
| **Avoidance: C1** | | | |
| Q01 | 0.78371 | 1.01195 | I prefer not to show a partner how I feel deep down. |
| Q03 | 4.65573 | 0.92730 | I am very comfortable being close to romantic partners. |
| *Q05* | *2.66541* | *1.28827* | *Just when my partner starts to get close to me I find myself pulling away.* |
| Q07 | 0.80563 | 0.97930 | I get uncomfortable when a romantic partner wants to be very close. |
| Q09 | 0.69756 | 0.96056 | I don't feel comfortable opening up to romantic partners. |
| Q11 | 0.24018 | 0.90108 | I want to get close to my partner, but I keep pulling back. |
| Q13 | 0.54060 | 0.83333 | I am nervous when partners get too close to me. |
| Q17 | 0.41101 | 0.83156 | I try to avoid getting too close to my partner. |
| Q21 | 1.39881 | 1.07866 | I find it difficult to allow myself to depend on romantic partners. |
| Q23 | -0.06296 | 0.78650 | I prefer not to be too close to romantic partners. |
| **Avoidance: C5** | | | |
| Q15 | 5.14500 | 1.02997 | I feel comfortable sharing my private thoughts and feelings with my partner. |
| Q19 | 1.87704 | 0.90586 | I find it relatively easy to get close to my partner. |
| Q25 | 3.03245 | 1.00122 | I tell my partner just about everything. |
| Q27 | 2.32169 | 0.80787 | I usually discuss my problems and concerns with my partner. |
| Q29 | 1.90582 | 0.95438 | I feel comfortable depending on romantic partners. |
| Q31 | 1.53331 | 0.88484 | I don't mind asking romantic partners for comfort, advice, or help. |
| Q33 | 2.08497 | 0.78218 | It helps to turn to my romantic partner in times of need. |
| Q35 | 0.33448 | 0.78453 | I turn to my partner for many things, including comfort and reassurance. |

**Table S4.** Intercepts and statements of all questions in the avoidance attachment style. The red items correspond to the reversed states of avoidance. The italicized items are the root nodes of the whole relationship network. The bolded items are the end nodes of the network.

| Q # | Intercept | Stdv | Questions / Items |
|---|---|---|---|
| **Anxiety: C2** | | | |
| *Q02* | *3.47722* | *1.30618* | *I worry about being abandoned.* |
| Q04 | 0.98243 | 0.96350 | I worry a lot about my relationships. |
| Q06 | 1.13748 | 1.04930 | I worry that romantic partners won't care about me as much as I care about them. |
| Q08 | 1.15097 | 1.00319 | I worry a fair amount about losing my partner. |
| Q10 | 0.79137 | 1.00087 | I often wish that my partner's feelings for me were as strong as my feelings for him/her. |
| Q14 | 0.60998 | 1.13592 | I worry about being alone. |
| Q22 | 5.36482 | 0.97469 | I do not often worry about being abandoned. |
| **Anxiety: C3** | | | |
| Q18 | 0.86593 | 1.04276 | I need a lot of reassurance that I am loved by my partner. |
| Q24 | 1.02465 | 1.01629 | If I can't get my partner to show interest in me, I get upset or angry. |
| Q28 | 0.43168 | 1.13675 | When I'm not involved in a relationship, I feel somewhat anxious and insecure. |
| Q30 | 1.07361 | 1.01026 | I get frustrated when my partner is not around as much as I would like. |
| Q32 | 0.63475 | 0.90112 | I get frustrated if romantic partners are not available when I need them. |
| **Q34** | **1.78689** | **1.09397** | **When romantic partners disapprove of me, I feel really bad about myself.** |
| **Q36** | **-0.04824** | **0.95068** | **I resent it when my partner spends time away from me.** |
| **Anxiety: C4** | | | |
| Q12 | 0.51011 | 1.03131 | I often want to merge completely with romantic partners, and this sometimes scares them away. |
| **Q16** | **0.22395** | **0.86281** | **My desire to be very close sometimes scares people away.** |
| Q20 | 0.55780 | 1.04815 | Sometimes I feel that I force my partners to show more feeling, more commitment. |
| Q26 | 0.88167 | 0.98057 | I find that my partner(s) don't want to get as close as I would like. |

**Table S5.** Intercepts and statements of all questions in the anxiety attachment style. The red texts correspond to the reversed states of anxiety. The italicized items are the root nodes of the whole relationship network. The bolded items are the end nodes of the network.

| From | To | Coefficient | | From | To | Coefficient |
|---|---|---|---|---|---|---|
| C1 | | | | C5 | | |
| ***Q05*** | Q01 | 0.12644 | | ***Q15*** | Q19 | 0.13026 |
| Q09 | Q01 | **0.42618** | | Q25 | Q19 | 0.10399 |
| Q17 | Q01 | **0.18693** | | Q27 | Q19 | 0.12152 |
| Q07 | **Q03** | **-0.22210** | | ***Q15*** | Q25 | **0.37907** |
| Q09 | **Q03** | -0.12882 | | ***Q15*** | Q27 | **0.18309** |
| Q13 | **Q03** | -0.13844 | | Q25 | Q27 | **0.36793** |
| Q23 | **Q03** | -0.16031 | | Q19 | **Q29** | 0.08801 |
| ***Q05*** | Q07 | **0.62769** | | Q31 | **Q29** | 0.14037 |
| Q07 | Q09 | **0.24156** | | Q33 | **Q29** | 0.11416 |
| Q11 | Q09 | **0.25704** | | Q35 | **Q29** | 0.11517 |
| Q13 | Q09 | **0.23551** | | ***Q15*** | Q31 | 0.09508 |
| ***Q05*** | Q11 | **0.43112** | | Q19 | Q31 | 0.12029 |
| Q13 | Q11 | **0.32562** | | Q27 | Q31 | **0.34872** |
| ***Q05*** | Q13 | **0.31203** | | Q25 | Q33 | 0.09396 |
| Q07 | Q13 | **0.47307** | | Q27 | Q33 | **0.17879** |
| ***Q05*** | Q17 | 0.13276 | | Q31 | Q33 | **0.29435** |
| Q07 | Q17 | 0.13106 | | Q27 | Q35 | 0.13884 |
| Q09 | Q17 | **0.18764** | | Q25 | Q35 | 0.11584 |
| Q11 | Q17 | 0.13307 | | Q31 | Q35 | **0.17217** |
| Q13 | Q17 | **0.21629** | | Q33 | Q35 | **0.37197** |
| Q09 | **Q21** | 0.14972 | | | | |
| Q11 | **Q21** | 0.11903 | | | | |
| Q13 | **Q21** | 0.12808 | | | | |
| Q17 | **Q21** | 0.12719 | | | | |
| Q23 | **Q21** | **0.19714** | | | | |
| Q01 | Q23 | 0.09244 | | | | |
| ***Q05*** | Q23 | 0.06054 | | | | |
| Q07 | Q23 | 0.16589 | | | | |
| Q09 | Q23 | 0.09900 | | | | |
| Q13 | Q23 | 0.10051 | | | | |
| Q17 | Q23 | **0.28728** | | | | |

**Table S6.** Influences between nodes within the 2 avoidance clusters C1 and C5. The italicized and the bolded items are respectively the root nodes and end nodes of their clusters. The red and dark blue items correspond to the positive and negative attachment styles respectively.

| From | To | Coefficient | | From | To | Coefficient | | From | To | Coefficient |
|---|---|---|---|---|---|---|---|---|---|---|
| **C2** | | | | **C3** | | | | **C4** | | |
| ***Q02*** | Q04 | **0.20662** | | ***Q18*** | Q24 | **0.31383** | | Q20 | Q12 | **0.19613** |
| Q06 | Q04 | **0.20561** | | Q30 | Q28 | **0.19883** | | ***Q26*** | Q12 | **0.31098** |
| Q08 | Q04 | **0.29901** | | ***Q18*** | Q30 | 0.14447 | | Q12 | **Q16** | **0.46561** |
| ***Q02*** | Q06 | **0.33685** | | Q24 | Q30 | **0.26274** | | Q20 | **Q16** | 0.09358 |
| Q08 | Q06 | **0.35162** | | ***Q18*** | Q32 | 0.08564 | | ***Q26*** | **Q16** | **0.19887** |
| ***Q02*** | Q08 | **0.58206** | | Q24 | Q32 | **0.22585** | | ***Q26*** | Q20 | **0.29496** |
| Q06 | **Q10** | **0.45768** | | Q30 | Q32 | **0.40243** | | | | |
| Q08 | **Q10** | **0.26974** | | ***Q18*** | **Q34** | **0.24224** | | | | |
| ***Q02*** | Q14 | **0.38212** | | Q24 | **Q34** | 0.15607 | | | | |
| Q04 | Q14 | **0.17722** | | Q28 | **Q34** | **0.17790** | | | | |
| Q08 | Q14 | **0.20448** | | Q24 | **Q36** | 0.10453 | | | | |
| ***Q02*** | **Q22** | **-0.41080** | | Q30 | **Q36** | **0.27007** | | | | |
| Q08 | **Q22** | **-0.19760** | | Q32 | **Q36** | 0.17126 | | | | |
| Q14 | **Q22** | -0.16130 | | | | | | | | |

**Table S7.** Influences between nodes within the 3 anxiety clusters C2, C3, and C4. The italicized and the bolded items are respectively the root nodes and end nodes of their clusters. The red and dark blue items correspond to the positive and negative attachment styles respectively.

| From | To | Coefficient | | From | To | Coefficient |
|---|---|---|---|---|---|---|
| **C2** | **C1** | | | **C1** | **C5** | |
| Q10 | Q11 | 0.13845 | | Q01 | ***Q15*** | **-0.22316** |
| **Q22** | Q23 | 0.11293 | | Q09 | ***Q15*** | **-0.39353** |
| | Sum abs | **0.25138** | | **Q03** | Q19 | **0.25354** |
| **C2** | **C3** | | | Q11 | Q19 | -0.16471 |
| ***Q02*** | ***Q18*** | **0.19283** | | Q01 | Q25 | **-0.19141** |
| Q04 | ***Q18*** | 0.14070 | | Q09 | Q25 | **-0.20569** |
| Q06 | ***Q18*** | 0.16411 | | Q01 | Q27 | -0.10530 |
| Q08 | ***Q18*** | 0.13531 | | Q09 | Q27 | -0.09305 |
| **Q10** | ***Q18*** | 0.12110 | | **Q03** | **Q29** | 0.11358 |
| Q14 | Q28 | **0.38309** | | **Q21** | **Q29** | **-0.31351** |
| Q14 | Q30 | 0.13297 | | **Q03** | Q31 | 0.10963 |
| Q08 | **Q36** | 0.10287 | | Q09 | Q31 | -0.10886 |
| **Q10** | **Q36** | 0.08132 | | Q23 | Q33 | -0.13276 |
| | Sum abs | **1.45430** | | | Sum abs | **2.40873** |
| **C2** | **C4** | | | **C3** | **C5** | |
| **Q10** | Q12 | **0.17397** | | ***Q18*** | Q35 | 0.13064 |
| Q14 | **Q16** | 0.08418 | | | Sum abs | 0.13064 |
| **Q10** | Q20 | 0.12310 | | **C4** | **C5** | |
| Q06 | ***Q26*** | 0.15982 | | ***Q26*** | Q19 | -0.09743 |
| **Q10** | ***Q26*** | **0.36447** | | | Sum abs | 0.09743 |
| | Sum abs | **0.90554** | | **C5** | **C3** | |
| **C3** | **C4** | | | **Q29** | Q28 | 0.10403 |
| ***Q18*** | Q20 | **0.32364** | | Q33 | Q32 | 0.10529 |
| | Sum abs | **0.32364** | | | Sum abs | **0.20932** |
| **C4** | **C3** | | | **C5** | **C1** | |
| Q20 | Q24 | **0.23781** | | ***Q15*** | **Q03** | **0.17482** |
| ***Q26*** | Q24 | 0.15089 | | | Sum abs | **0.17482** |
| Q20 | Q30 | 0.13498 | | | | |
| Q20 | **Q36** | 0.08301 | | | | |
| | Sum abs | **0.60669** | | | | |

**Table S8.** Influences between nodes from the different clusters (red, all positive; dark blue, all negative; purple, both). Red and dark blue items correspond to positive and negative attachment styles respectively. Pink highlighted cells correspond to positive coefficients between positive and negative attachment states.

| **A** | | | | | | | | | | | |
|---|---|---|---|---|---|---|---|---|---|---|---|
| **Q02** | 1 | 1 | 1 | 1 | 1 | 1 | 1 | 1 | | | |
| **Q08** | | | 0.5821 | 0.5821 | 0.5821 | 0.5821 | 0.5821 | | | | |
| **Q06** | | 0.3369 | 0.3369 | | | | | | | | |
| **Q04** | 0.2066 | 0.2056 | 0.2056 | 0.2990 | | | | | | | |
| **Q10** | | | | | | | | | | | |
| **Q14** | 0.1772 | 0.1772 | 0.1772 | 0.1772 | 0.2045 | | 0.3821 | | | | |
| **Q22** | -0.1613 | -0.1613 | -0.1613 | -0.1613 | -0.1613 | -0.1976 | -0.1976 | -0.4108 | | | |
| **Q23** | 0.1129 | 0.1129 | 0.1129 | 0.1129 | 0.1129 | 0.1129 | 0.1129 | 0.1129 | | | |
| **Product** | -0.0007 | -0.0002 | -0.0001 | -0.0006 | -0.0022 | -0.0130 | -0.0070 | -0.0464 | **Sum** | -0.0701 | |
| | | | | | | | | | | | |
| **B** | | | | | | | | | | | |
| **Q05** | 1 | 1 | 1 | 1 | 1 | 1 | 1 | 1 | | | |
| **Q07** | | | | | | 0.6277 | 0.6277 | | | | |
| **Q13** | | | 0.3120 | | 0.3120 | | | | | | |
| **Q11** | 0.4311 | | | 0.4311 | | | | | | | |
| **Q09** | | | | 0.2570 | 0.2355 | 0.2416 | | | | | |
| **Q17** | 0.1331 | 0.1328 | 0.2163 | | | | | | | | |
| **Q01** | | | | | | | | | | | |
| **Q23** | 0.2873 | 0.2873 | 0.2873 | 0.0990 | 0.0990 | 0.0990 | 0.1659 | 0.0605 | | | |
| **Product** | 0.0165 | 0.0382 | 0.0194 | 0.0110 | 0.0073 | 0.0150 | 0.1041 | 0.0605 | **Sum** | 0.2719 | |

**Table S9.** Indirect influences from the root nodes Q02 and Q05 to Q23. Indirect or overall influences can be measured as a sum of coefficient products along the paths. Each column represents one of the paths from Q02 to Q23 (A) or from Q05 to Q23 (B). Each value along a column is a coefficient from the previous valued node to the current node of the same row with the value. For example, in the first column/path of Q05 to Q23, the values 0.4311, 0.1331, and 0.2873 are coefficients of the arcs (Q05, Q11), (Q11, Q17), and (Q17, Q23) respectively.  Q02 has an overall negative influence on Q23 due to the negative influence of Q22.

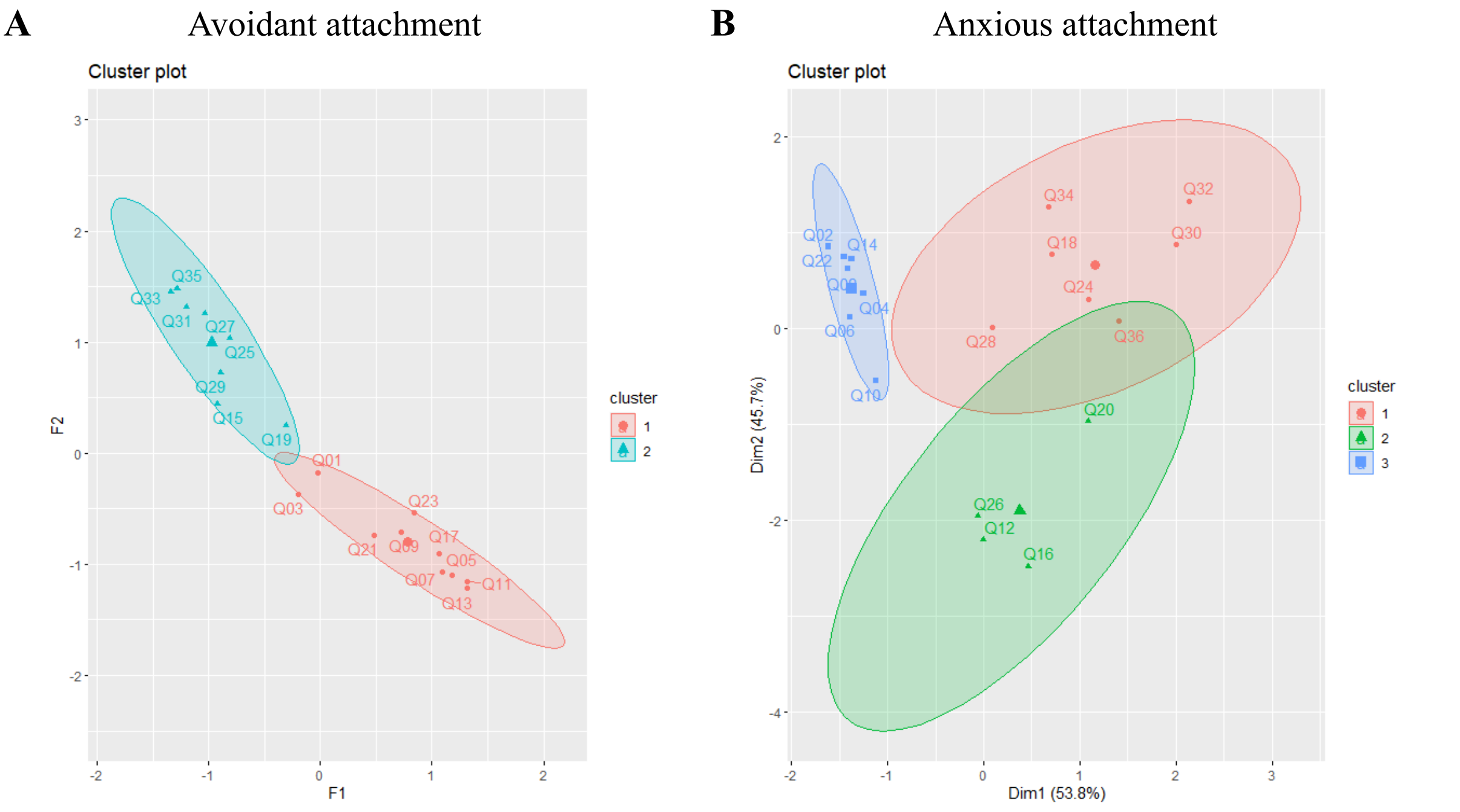

**Figure S5.** Cluster plots of avoidant and anxious attachment items from Wei et al. The 18 avoidance items are plotted on the avoidance factor 1 and factor 2 (A). As there are 3 anxiety factors, 18 anxiety items are plotted on PC1 and PC2 of the 3 factors (B). Note that choices of color here between the two plots do not portray any specific features of the data.

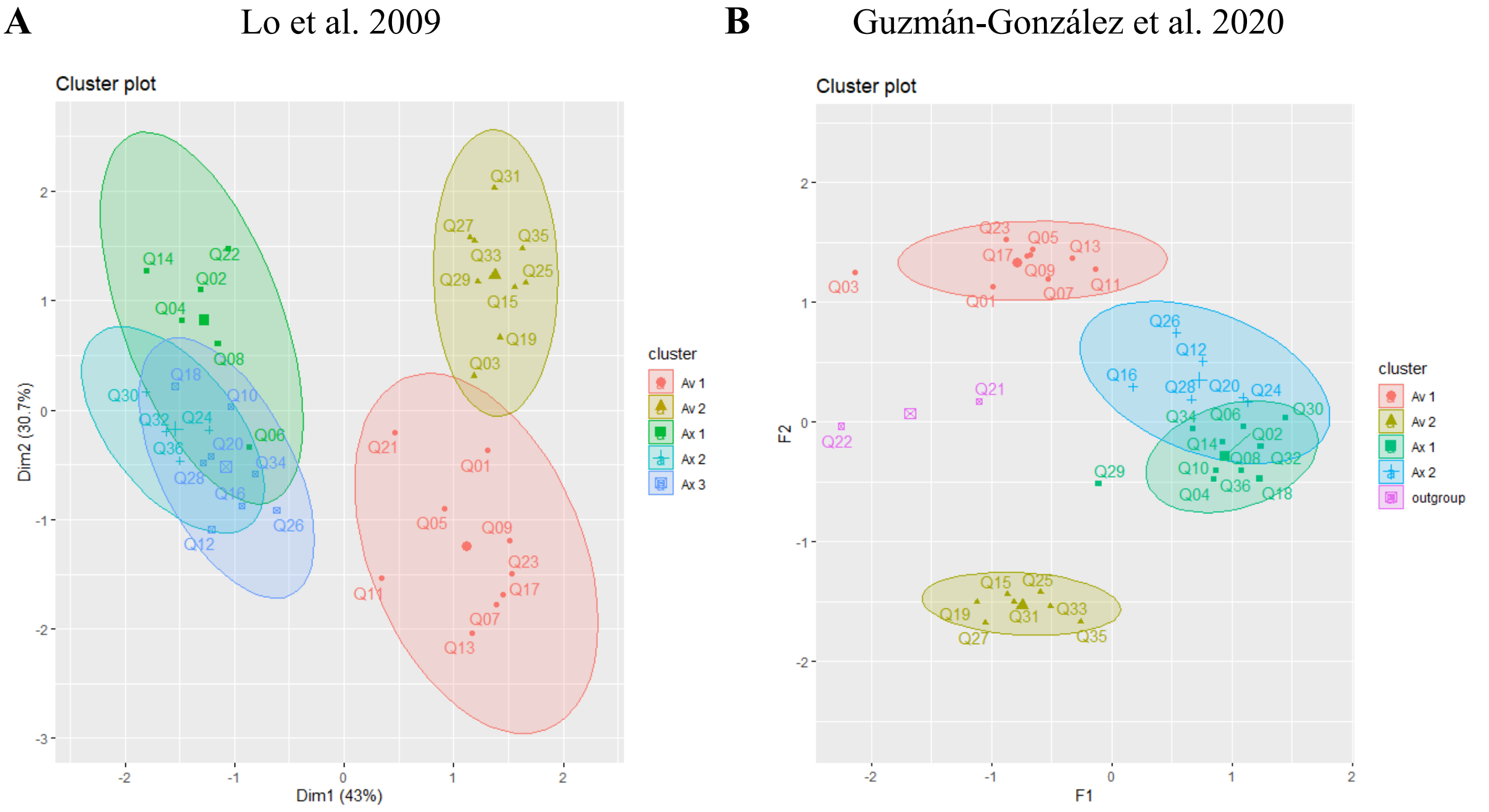

**Figure S6.** Cluster plots of items from Lo et al. and Guzmán-González et al. All 36 items together are plotted on the PC1 and PC2 of 4 factors from Lo et al. (A) and on the factor 1 and factor 2 from Guzmán-González et al. (B). Red and yellow denote clusters of avoidant attachment specifically associated with our C1 and C5, whereas green to blue color range denotes clusters of anxious attachment with characteristics gradient from C2 to C4 respectively. Violet color, only in (B), represents the outgroup**.**

**A** **McWilliams & Fried 2019 (RSQ)**

| | **1** | **2** | **3** | **4** | **5** | **6** | **7** | **8** | **9** |
|---|---|---|---|---|---|---|---|---|---|
| **1** | | 0.523 | 0.393 | 0.032 | 0.139 | 0.104 | | -0.01 | |
| **2** | | | 0.074 | 0.15 | 0.043 | 0.042 | | -0.018 | |
| **3** | | | | 0.341 | 0.017 | 0.019 | -0.033 | -0.005 | 0.033 |
| **4** | | | | | | | | | 0.080 |
| **5** | | | | | | 0.426 | -0.020 | | 0.068 |
| **6** | | | | | | | 0.003 | | 0.062 |
| **7** | | | | | | | | 0.434 | 0.305 |
| **8** | | | | | | | | | 0.424 |

**B** **Current study (ECL)**

| | **27** | **25** | **33** | **29** | **1** | **9** | **2** | **6** | **18** |
|---|---|---|---|---|---|---|---|---|---|
| **27** | | 0.368 | 0.179 | 0 | 0.105 | 0.093 | | 0 | |
| **25** | | | 0.094 | 0 | 0.191 | 0.206 | | 0 | |
| **33** | | | | 0.114 | 0 | 0 | 0 | 0 | 0 |
| **29** | | | | | | | | | 0 |
| **1** | | | | | | 0.426 | 0 | | 0 |
| **9** | | | | | | | 0 | | 0 |
| **2** | | | | | | | | 0.337 | 0.193 |
| **6** | | | | | | | | | 0.164 |

**C**

| | |
|---|---|
| **df** | 24 |
| **r** | 0.823 |
| **$r^2$** | 0.677 |
| **t** | 7.094 |
| ***p*** | 2.48E-07 |

**D**

| | |
|---|---|
| **df** | 10 |
| **r** | 0.626 |
| **$r^2$** | 0.392 |
| **t** | 2.54 |
| ***p*** | 0.0294 |

**Table S10.** Correlation between network weights of our study and McWilliams and Fried. Partial correlation from the network study (A) and causal influence from our study (B) are listed. Light yellow indicates items in RSQ that have similar but not exact statements with their ECR counterparts. The correlation between the 26 edge pairs shown in either of the 2 studies (C) and the 12 pairs shown in both studies (D).